\documentclass[journal,final]{IEEEtran}

\IEEEoverridecommandlockouts

\usepackage{xspace,amsmath,amssymb,epsfig,syntonly,color, cite}

\newtheorem{lemma}{\hspace{-11pt}\bf Lemma}
\newtheorem{algorithm}{\hspace{-11pt}\bf Algorithm}

\newtheorem{proposition}{\hspace{-11pt}\bf Proposition}
\newtheorem{theorem}{\hspace{-11pt}\bf Theorem}

\newtheorem{remark}{\hspace{-11pt}\bf Remark}

\begin{document}
\title{ Towards the Asymptotic Sum Capacity of the MIMO Cellular Two-Way Relay Channel }

\author{
   {Zhaoxi~Fang},
   {Xiaojun~Yuan}, ~\IEEEmembership{Member,~IEEE}
   {and Xin~Wang}, ~\IEEEmembership{Senior Member,~IEEE}\\
    \thanks{

     Work in this paper was supported by the China Recruitment Program of Global Young Experts, the General Research Funds Project No. 418712 established under the University Grant Committee of the Hong Kong Special Administrative Region, China, the National Key Technology R\&D Program of China under Grant No. 2013BAK09B03, and the Natural Science Foundation of Ningbo City under Grant No. 2013A610121.

    Z. Fang is with the Faculty of Electronic and Information Engineering,
Zhejiang Wanli University, Ningbo, China.

 X. Yuan is with the
Institute of Network Coding, Dept. of Information Engineering, The
Chinese University of Hong Kong, e-mail:
xjyuan{\rm\char64}inc.cuhk.edu.hk.

X. Wang is with Dept. of Communication Science and Engineering,
Fudan University, Shanghai, China, email:
xwang11{\rm\char64}fudan.edu.cn.

The corresponding author is X. Yuan (e-mail:
xjyuan{\rm\char64}inc.cuhk.edu.hk).
} }

\maketitle 

\vspace{-1 in}

\begin{abstract}
In this paper, we consider the transceiver and relay design for the multiple-input multiple-output (MIMO) cellular two-way relay channel (cTWRC), where a multi-antenna base station (BS) exchanges information with multiple multi-antenna mobile stations via a multi-antenna relay station (RS). We propose a novel  two-way relaying scheme to approach the sum  capacity of the MIMO cTWRC. A key contribution of this work is a new non-linear lattice-based precoding technique to pre-compensate the inter-stream interference, so as to achieve efficient interference-free lattice decoding at the relay. We derive sufficient conditions for the proposed scheme to asymptotically achieve the sum capacity of the MIMO cTWRC in the high signal-to-noise ratio (SNR) regime. To fully exploit the potential of the proposed scheme, we also investigate the optimal power allocation  at the BS and the RS to maximize the weighted sum-rate of the MIMO cTWRC in the general SNR regime. It is shown that the problem can be formulated as a monotonic program, and a polyblock outer approximation algorithm is developed to find the globally optimal solution with guaranteed convergence. We demonstrate by numerical results that the proposed scheme significantly outperforms the existing schemes and closely approaches the sum capacity of the MIMO cTWRC in the high SNR regime.
\end{abstract}

\begin{keywords}
Cellular two-way relay channel, nested lattice coding, lattice precoding, monotonic optimization.
\end{keywords}


\section{Introduction}

Two-way communications can be dated back to Shannon \cite{Shannon61} and has been rediscovered as an efficient method to mitigate the loss of spectral efficiency in conventional half-duplex one-way relaying \cite{Zhang06,Ran07,Pop07,Naz11,Nam10,Yang13,Yuan13,Yang11}. Tremendous progress has been made for efficient communications over the two-way relay channel (TWRC), in which two users want to exchange information via the help of a single relay. The main idea, termed physical-layer network coding (PNC), is to allow the two users to communicate with the relay simultaneously, and to allow each user to decode the message from the other user by exploiting the knowledge of the self-message. It was shown in \cite{Nam10} that PNC with nested lattice coding can achieve the cut-set outer bound of the single-input single-output  Gaussian TWRC within $1/2$ bit. Later, the authors in \cite{Yang13,Yuan13,Yang11} studied the multiple-input multiple-output (MIMO) TWRC, where both the users and the relay are equipped with multiple antennas. It was shown that near-capacity performance can be achieved in the MIMO TWRC by using nested lattice coding aided PNC.

PNC design for more sophisticated relay networks has recently attracted much research interest. In this regard, the authors in \cite{Gun13}  generalized the TWRC model to the multiway relay channel (mRC), in which a relay simultaneously serves multiple clusters of users, and each user in a cluster wants to multicast its message to all the other users in the same cluster. Several special cases of the mRC have been studied in the literature. For example, the authors in \cite{Che09,Tao12,Zha12,Fang13} considered the multi-pair MIMO TWRC, which is a special case of the mRC with each cluster consisting of two users; also, the Y channel proposed in \cite{Lee10} is a special case of the mRC with only one cluster. Various relaying protocols, including amplify-and-forward (AF), decode-and-forward (DF), compress-and-forward, and their mixtures, were investigated for these relay networks.

In this paper, we investigate efficient PNC design for another important two-way relaying model, termed the cellular TWRC (cTWRC), where multiple users in a cellular network want to exchange information with a multiple-antenna base station (BS) via the help of a multiple-antenna relay station (RS). It has been shown that two-way relaying has the potential to increase network throughput and extend the coverage, and is a promising technique for future cellular systems \cite{Bhat12}. For this reason,  a  MIMO cTWRC model was previously studied in \cite{Ding11,Sun12,Chi12,Gan13,YangHJ12,Wan13}. In this model, the information exchange is realized using a two-phase protocol: In the first phase both the BS and the users transmit signals to the BS; in the second phase, the relay broadcasts signals to the BS and the users.  In \cite{Ding11}, linear precoding was applied at the BS to align the signals impinging upon the relay in such a way that each signal stream of the BS is aligned to the direction of the user's signal stream to be exchanged with. In \cite{Sun12} and \cite{Chi12}, linear precoding was applied at both the BS and the relay, and iterative algorithms were proposed to optimize the corresponding precoders based on various design criteria, such as   sum-rate maximization or max-min  signal-to-interference-plus-noise ratio (SINR). However, all these approaches were based on AF relaying which generally suffers from the noise propagation, as well as from the power inefficiency caused by transmitting analogue (instead of algebraic) superposition  of the BS and user signals at the relay.

To avoid the aforementioned disadvantages, the authors in \cite{YangHJ12} proposed a DF-based relaying scheme for the MIMO cTWRC involving linear precoding and nested lattice coding at the BS and users, and dirty-paper coding at the RS. It was shown that the achievable sum-rate of this scheme is much higher than the AF based schemes in \cite{Ding11,Sun12,Chi12}, and this scheme can achieve the cut-set outer bound of the MIMO cTWRC if only the second phase is concerned. However, the scheme in \cite{YangHJ12} can perform far away from  the capacity of the MIMO cTWRC, especially for a relatively large MIMO setup. This performance gap is largely due to the following fact: In the first phase, linear precoding can be applied only at the BS (as the users cannot cooperate), whereas the BS precoder alone fails to provide enough freedom to align the signals efficiently for interference-free PNC decoding at the relay.

In this paper, we propose a novel nested-lattice-coding aided PNC scheme to approach the sum capacity of the MIMO cTWRC. Compared to  \cite{YangHJ12}, a major difference and contribution of this work is a new non-linear precoding technique, called lattice precoding, employed in the first-phase system design. We show that, together with linear precoding, nested lattice coding, and successive interference cancellation (SIC), the proposed lattice precoder at the BS efficiently pre-compensates for the inter-stream interference (ISI) seen at the relay, such that interference-free lattice decoding can be performed at the relay. We derive the achievable rates of the proposed scheme, and establish sufficient conditions for the proposed scheme to asymptotically achieve the sum capacity of the MIMO cTWRC  in the high SNR regime. Furthermore, we formulate a weighted sum-rate maximization problem  for the proposed scheme to optimize the power allocation of the nodes in the network and show that this non-convex problem is solvable using monotonic programming (MP) \cite{Tuy00}. An efficient polyblock outer approximation algorithm is developed to find the optimal power allocation. Numerical results demonstrate that the proposed scheme significantly outperforms its existing alternatives and closely approaches the capacity of the MIMO cTWRC at high SNR.

The rest of this paper is organized as follows. Section II describes the MIMO cTWRC model. Section III presents the proposed encoding and decoding scheme, while Section IV analyzes the achievable sum-rate of the proposed scheme. Section V investigates the optimal power allocation problem. Section VI discusses the extension of the proposed scheme to MIMO cTWRCs with general antenna setups. The proposed scheme is tested and compared with the existing schemes in Section VII, followed by the conclusions in Section VIII.

\textit{Notation:} The following notation is used throughout this paper. Boldface letters
denote vectors or matrices. $(\cdot)^T$ denotes
transpose.  The $(i,j)$-th element of a matrix $\boldsymbol{A}$ is denoted by $a(i,j)$.  $\mathbb{R}^{K\times M}$ and $\mathbb{C}^{K\times M}$ denote the
$K$-by-$M$ dimensional real and complex space, respectively. $\| \cdot \|_{\text{F}}$ denotes
the Frobenius norm. $\text{tr}(\cdot)$ denotes the trace operation; $\text{diag}(a_1, \ldots, a_N)$ denotes a
diagonal matrix with diagonal elements $(a_1, \ldots, a_N)$, and $(\boldsymbol{A})_{\text{diag}}$ denotes the diagonal matrix specified by the diagonal of matrix $\boldsymbol{A}$; $\boldsymbol{I}_K$ denotes a $K \times K $ identity matrix; $\boldsymbol{e}_i$ denotes a unit vector with the only non-zero element in the $i$-th entry; $[x]^+$ denotes $\max(x,0)$.

\section{System Model}


\begin{figure}[t]
\centering
\includegraphics[width=3.3in]{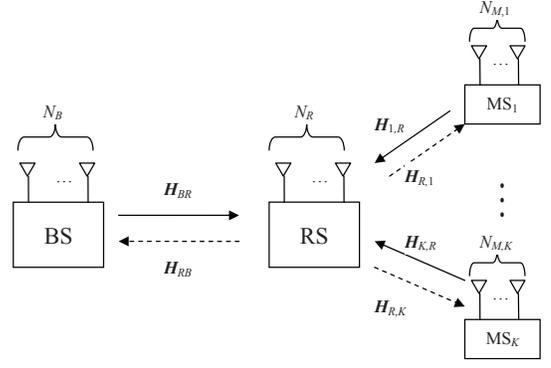}
\caption{ A  MIMO cTWRC with $K$ mobile stations. }
 \label{fig1}
\end{figure}

We consider a MIMO cTWRC where a BS communicates with $K$ mobile stations (MSs) via a single relay station, as shown in Fig. 1. There is no direct link between the BS and the MSs. The BS, the RS, and the $k$-th MS are equipped with $N_B$, $N_R$, and $N_{M,k}$ antennas, respectively.  We consider quasi-static flat-fading channels where the channel coefficients keep unchanged in the duration of a transmission frame, denoted by $T$.  All the nodes are half-duplex and the bidirectional transmission takes place in two phases. For presentation clarity, we consider single-antenna MSs, i.e., $N_{M,k}=1$, for $k=1,\ldots,K$, and assume $N_B=N_R=K$. The extension to a general antenna setup will be discussed in Section VI. Each MS exchanges one data stream with the BS, and there are $2K$ data streams in total. The channel matrix from  the BS to the RS is denoted by $\boldsymbol{H}_{BR} \in \mathbb{C} ^{K \times K}$, and the channel vector from  the $k$-th MS to the RS  by $\boldsymbol{h}_{k,R} \in \mathbb{C} ^{K \times 1}$. $\boldsymbol{H}_{RB} \in \mathbb{C} ^{K \times K}$ and $\boldsymbol{h}_{R,k} \in \mathbb{C} ^{1 \times K} $ are the corresponding channel matrix/vector for the reverse links. Following the convention (e.g., in \cite{Sun12} and \cite{YangHJ12}), we assume that all the nodes have global channel state information of all links.

The transmission protocol is described as follows. In the first phase, the BS and all the $K$ MSs transmit to the RS simultaneously.  Let $\boldsymbol{X}_{B} \in \mathbb{C}^{K \times T}$ and $\boldsymbol{x}_{M,k} \in \mathbb{C}^{1\times T}$ denote the transmit signal at the BS and the $k$-th MS, respectively. The received signal at the RS is given by
\begin{equation}\label{eq.yR}
\boldsymbol{Y}_R = \boldsymbol{H}_{BR} \boldsymbol{X}_B  +
\boldsymbol{H}_{MR} \boldsymbol{X}_M  + \boldsymbol{\Psi}_R ,
\end{equation}
where $\boldsymbol{H}_{MR} := [\boldsymbol{h}_{1,R},\ldots,\boldsymbol{h}_{K,R}] \in \mathbb{C}^{K \times K}$,
$\boldsymbol{X}_M  := [\boldsymbol{x}_{M,1}^T ,\ldots,\boldsymbol{x}_{M,K}^T ]^T \in \mathbb{C}^{K \times T}$, and $\boldsymbol{\Psi}_R \in \mathbb{C}^{K \times T}$ denotes the additive white Gaussian noise (AWGN) at the RS. It is assumed that each element in $\boldsymbol{\Psi}_R $ is independent and identically distributed (i.i.d.) with zero mean and a variance of $\sigma^2$. The maximum transmit powers at the BS and the $k$-th MS are $P_B$ and $P_{M,k}$ respectively, i.e., $\frac{1}{T} \text{tr}(\boldsymbol{X}_B \boldsymbol{X}_B^H) \le P_B$, and $\frac{1}{T} \|\boldsymbol{x}_{M,k}\|_{\text{F}}^2 \le P_{M,k}$.

Upon receiving $\boldsymbol{Y}_R $, the transmit signal at the RS is regenerated as $\boldsymbol{X}_{R}  = g_{R}(\boldsymbol{Y}_{R} ) \in \mathbb{C}^{K \times T}$, where $g_{R}(\cdot)$ denotes the RS decoding and re-encoding function. The transmit power of the RS is constrained as $\frac{1}{T} \text{tr}(\boldsymbol{X}_R \boldsymbol{X}_R^H) \le P_R$, where $P_R$ is the power budget at the relay.

In the second phase, the RS broadcasts  $\boldsymbol{X}_{R} $  to the BS and the $K$ MSs. The received signals at the BS and the $k$-th MS are given by
\begin{equation}\label{eq.yb}
\boldsymbol{Y}_{B}  = \boldsymbol{H}_{RB}\boldsymbol{X}_{R}  + \boldsymbol{\Psi}_{B},
\end{equation}
\begin{equation}\label{eq.ymk}
\boldsymbol{y}_{M,k}  = \boldsymbol{h}_{R,k} \boldsymbol{X}_{R}  + \boldsymbol{\psi}_{M,k} ,
\end{equation}
where $\boldsymbol{\Psi}_B $ and $\boldsymbol{\psi}_{M,k} $ are the AWGN at the BS and the $k$-th MS, respectively.  With $\boldsymbol{Y}_{M} = [\boldsymbol{y}_{M,1}^T, \ldots,\boldsymbol{y}_{M,K}^T]^T$, $\boldsymbol{H}_{RM}= [\boldsymbol{h}_{R,1}^T, \ldots,\boldsymbol{h}_{R,K}^T]^T$, and $\boldsymbol{\Psi}_{M} = [\boldsymbol{\psi}_{M,1}^T, \ldots,\boldsymbol{\psi}_{M,K}^T]^T$, we have
\begin{equation}\label{eq.ym}
\boldsymbol{Y}_{M}  = \boldsymbol{H}_{RM} \boldsymbol{X}_{R}  + \boldsymbol{\Psi}_{M} .
\end{equation}

With the knowledge of the self-message $\boldsymbol{X}_B$, the BS estimates all the MS messages from the received signal $\boldsymbol{Y}_{B} $. Meanwhile, for each  $k \in \{1, \ldots, K\}$, the $k$-th MS  decodes the intended private message of the BS from $\boldsymbol{y}_{M,k} $ with the knowledge of $\boldsymbol{x}_{M,k} $.

For the MIMO cTWRC, let $R_{B,k}$ be the transmission rate from the BS to the $k$-th MS, and $R_{M,k}$ be the transmission rate from the $k$-th MS to the BS. A rate tuple $(R_{B,1}, \ldots, R_{B,K}, R_{M,1}, \ldots, R_{M,K})$ is said to be achievable if there exist transmit encoding functions, MIMO processing functions, and receive decoding functions  at the BS, RS, and MSs such that the decoding error probabilities tend to zero as the codeword length $T \to \infty$. From the cut-set theorem, two sum-rate outer bounds of the MIMO cTWRC are given by \cite{YangHJ12}
\begin{subequations}\label{eq.cutset1}
\begin{align}
\sum_{k=1}^{K} R_{B,k} & \le \min \left( \frac{1}{2} \log |\boldsymbol{I}_K + \boldsymbol{H}_{BR} \boldsymbol{Q}_B \boldsymbol{H}_{BR}^H |, \right. \nonumber  \\
& \left. \quad \frac{1}{2} \log |\boldsymbol{I}_K + \boldsymbol{H}_{RM} \boldsymbol{Q}_R \boldsymbol{H}_{RM}^H | \right), \\
\sum_{k=1}^{K} R_{M,k} & \le \min \left( \frac{1}{2} \log |\boldsymbol{I}_K + \boldsymbol{H}_{MR} \boldsymbol{Q}_M \boldsymbol{H}_{MR}^H | ,  \right. \nonumber  \\
&  \left. \quad ~  \frac{1}{2} \log |\boldsymbol{I}_K + \boldsymbol{H}_{RB} \boldsymbol{Q}_R \boldsymbol{H}_{RB}^H |\right),
\end{align}
\end{subequations}
where $\boldsymbol{Q}_{\cal{S}} = \frac{1}{T} E(\boldsymbol{X}_{\cal{S}} \boldsymbol{X}_{\cal{S}}^H), {\cal{S}} \in \{B,R,M\}$, are the corresponding signaling covariance matrices. These bounds will be used as a benchmark of the system design for the MIMO cTWRC.

\section{Proposed Two-way Relaying Scheme}

In this section, we propose a novel two-phase two-way relaying scheme to approach the sum-rate capacity of the MIMO cTWRC. The key novelty of our scheme, compared to  \cite{YangHJ12}, is that lattice precoding and random dithering are employed in the first phase to pre-compensate for the inter-stream interference.  Building on this theme, encoding and decoding operations at the BS, the RS, and the MSs are carefully designed to enable efficient interference-free PNC decoding at the relay, even with the restriction of non-cooperation among MSs.

\subsection{Channel Triangularization}
To start with, we describe a linear precoding technique to triangularize the channel matrices involved in the two transmission phases, following the approach in \cite{YangHJ12}.

Consider the first phase. Let the QR decomposition of $\boldsymbol{H}_{MR}$ be
\begin{equation}
\boldsymbol{H}_{MR} = \boldsymbol{Q}_{MR} \boldsymbol{R}_{MR},
\end{equation}
where $\boldsymbol{Q}_{MR}$ is a unitary matrix and $ \boldsymbol{R}_{MR}$ is an upper-triangular matrix.  Further let the RQ decomposition of $\boldsymbol{Q}_{MR}^H \boldsymbol{H}_{BR}$ be
\begin{equation}
\boldsymbol{Q}_{MR}^H \boldsymbol{H}_{BR} = \boldsymbol{R}_{BR} \boldsymbol{Q}_{BR},
\end{equation}
where $\boldsymbol{Q}_{BR}$ is a unitary matrix and $ \boldsymbol{R}_{BR}$ is an upper-triangular matrix. By multiplying $\boldsymbol{Q}_{MR}^H$ to the RS received signal $\boldsymbol{Y}_R $, we obtain
 \begin{equation}\label{eq.yR2}
\tilde{\boldsymbol{Y}}_R  = \boldsymbol{Q}_{MR}^H \boldsymbol{Y}_R = \boldsymbol{R}_{BR} \tilde{\boldsymbol{X}}_B  +  \boldsymbol{R}_{MR} \boldsymbol{X}_M  + \tilde{\boldsymbol{\Psi}}_R,
\end{equation}
where $\tilde{\boldsymbol{X}}_B  =  \boldsymbol{Q}_{BR}\boldsymbol{X}_B $, and $\tilde{\boldsymbol{\Psi}}_R  = \boldsymbol{Q}_{MR}^H \boldsymbol{\Psi}_R $.

Let $\boldsymbol{s}_{B,k} \in \mathbb{C}^{1 \times T}$ be the coded vector of the BS transmitted to the $k$-th MS. The transmit signal of the BS is linearly precoded as $\boldsymbol{X}_{B} = \boldsymbol{Q}_{BR}^H \boldsymbol{S}_{B}$, where $\boldsymbol{S}_{B} =[\boldsymbol{s}_{B,1}^T, \ldots, \boldsymbol{s}_{B,K}^T]^T \in \mathbb{C}^{K \times T}$ is the codeword matrix. The transmit signal of the $k$-th MS is directly generated as $\boldsymbol{x}_{M,k} = \boldsymbol{s}_{M,k}$, where $\boldsymbol{s}_{M,k} \in \mathbb{C}^{1 \times T}$ denotes the coded vector of the $k$-th MS. With such a linear precoding, the RS obtains
 \begin{equation}\label{eq.yR3}
\tilde{\boldsymbol{Y}}_R  = \boldsymbol{R}_{BR} \boldsymbol{S}_B  +  \boldsymbol{R}_{MR} \boldsymbol{S}_M  + \tilde{\boldsymbol{\Psi}}_R,
\end{equation}
where $\boldsymbol{S}_M = [\boldsymbol{s}_{M,1}^T , \ldots, \boldsymbol{s}_{M,K}^T]^T \in \mathbb{C}^{K \times T}$. Correspondingly, the power constraints at the BS and the $k$-th MS can be equivalently written as  $\frac{1}{T} \text{tr}(\boldsymbol{S}_B \boldsymbol{S}_B^H) \le P_B$, and $\frac{1}{T} \|\boldsymbol{s}_{M,k}\|^2 \le P_{M,k}, k=1,\ldots,K$.

The channels from the RS to the BS and MSs can be triangularized in a similar way. Let $\boldsymbol{\Phi}$ be an arbitrary  permutation matrix.  We will see in Section III-{\it E} that $\boldsymbol{\Phi}$ specifies the DPC re-encoding order at the relay. Let the LQ decomposition of  the re-ordered channel matrix $\boldsymbol{\Phi} \boldsymbol{H}_{RM}$ be
\begin{equation}\label{eq.LQ}
\boldsymbol{\Phi} \boldsymbol{H}_{RM} = \boldsymbol{L}_{RM} \boldsymbol{Q}_{RM},
\end{equation}
where  $\boldsymbol{Q}_{RM}$ is unitary  and $ \boldsymbol{L}_{RM}$ is lower-triangular.

The transmit signal of the RS is precoded as
\begin{equation}\label{eq.xr}
\boldsymbol{X}_{R}  = \boldsymbol{Q}_{RM}^H \boldsymbol{X}_{R,DPC},
\end{equation}
where $\boldsymbol{X}_{R,DPC} \in \mathbb{C}^{K \times T}$ is the DPC codeword matrix to be elaborated in Subsection {\it E}. As $\boldsymbol{Q}_{RM}$ is unitary, the power constraint of the relay can be equivalently written as $\frac{1}{T} \text{tr}(\boldsymbol{X}_{R,DPC} \boldsymbol{X}_{R,DPC}^H) \le P_R$.

The permuted received signal $\tilde{\boldsymbol{Y}}_{M}  = \boldsymbol{\Phi}\boldsymbol{Y}_{M} $  at all $K$ MSs can be expressed as
\begin{equation}\label{eq.ym2}
\tilde{\boldsymbol{Y}}_{M}  = \boldsymbol{\Phi}\boldsymbol{H}_{RM}\boldsymbol{X}_{R}  +  \boldsymbol{\Phi}\boldsymbol{\Psi}_{M}= \boldsymbol{L}_{RM}\boldsymbol{X}_{R,DPC}  + \tilde{\boldsymbol{\Psi}}_{M},
\end{equation}
where $\tilde{\boldsymbol{\Psi}}_{M}  = \boldsymbol{\Phi}\boldsymbol{\Psi}_{M} $ is still an AWGN matrix. Note that there is only one non-zero entry in each row and column of the permutation matrix $\boldsymbol{\Phi}$. For a given $\boldsymbol{\Phi}$,   $\boldsymbol{\Phi} {\boldsymbol{Y}}_{M} $ gives a matrix with reordered rows of $\boldsymbol{Y}_{M}$. This operation doesn't involve any joint processing of the signals from different MSs. Therefore, the permutation operation $\tilde{\boldsymbol{Y}}_{M} = \boldsymbol{\Phi} {\boldsymbol{Y}}_{M} $ is possible, even when the MSs can not cooperate. Based on the signal model in (\ref{eq.ym2}), we will show in Section III-{\it F} that the achievable rates of the RS-MS link depend on the diagonal of $ \boldsymbol{L}_{RM}$.  On the other hand,  it follows from (\ref{eq.LQ}) that the lower-triangular matrix $ \boldsymbol{L}_{RM}$ varies with the choice of the permutation matrix $\boldsymbol{\Phi}$. Hence, the sum-rate performance of the proposed scheme can be improved by searching the optimal permutation matrix $\boldsymbol{\Phi}$.

Now consider the received signal at the BS. Let the QL decomposition of $\boldsymbol{H}_{RB} \boldsymbol{Q}_{RM}^H $ be
\begin{equation}
\boldsymbol{H}_{RB} \boldsymbol{Q}_{RM}^H = \boldsymbol{Q}_{RB} \boldsymbol{L}_{RB}.
\end{equation}
where  $\boldsymbol{Q}_{RB}$ is unitary  and $ \boldsymbol{L}_{RB}$ is lower-triangular. By multiplying $\boldsymbol{Q}_{RB}^H$ to the received signal $\boldsymbol{Y}_{B}$ in (\ref{eq.yb}), the BS obtains
\begin{equation}\label{eq.yb2}
\tilde{\boldsymbol{Y}}_{B}  = \boldsymbol{Q}_{RB}^H \boldsymbol{Y}_{B}  =\boldsymbol{L}_{RB}\boldsymbol{X}_{R,DPC}  + \tilde{\boldsymbol{\Psi}}_{B},
\end{equation}
where $\tilde{\boldsymbol{\Psi}}_{B}= \boldsymbol{Q}_{RB}^H \boldsymbol{\Psi}_{B}$. In the above channel triangularization, only unitary transforms are involved, implying that the new signal model in (\ref{eq.yR3}), (\ref{eq.ym2}) and  (\ref{eq.yb2}) has the same capacity as the original MIMO cTWRC. Therefore, we henceforth focus on the signaling design for the equivalent MIMO cTWRC given by  (\ref{eq.yR3}), (\ref{eq.ym2}) and  (\ref{eq.yb2}).

\subsection{Partitions of Upper-Triangular Matrices $\boldsymbol{R}_{BR}$ and $\boldsymbol{R}_{MR}$}

In the following, we describe the main ideas behind our novel lattice precoding and decoding scheme, based on channel models given in (\ref{eq.yR3}), (\ref{eq.ym2}) and  (\ref{eq.yb2}). We first consider the phase-1 channel model in (\ref{eq.yR3}).  From (\ref{eq.yR3}),  we can see that $ \boldsymbol{R}_{BR}$ and $\boldsymbol{R}_{MR}$ are both upper-triangular matrices. Hence, ISI generally exists in relay decoding. Our objective is to cancel the ISI by combining lattice precoding at the BS and successive interference cancellation at the relay. To this end, we rewrite $\boldsymbol{R}_{BR}$ and $\boldsymbol{R}_{MR}$ as $\boldsymbol{R}_{BR} = \boldsymbol{R}_{BR}^{'} + (\boldsymbol{R}_{BR} - \boldsymbol{R}_{BR}^{'})$ and $\boldsymbol{R}_{MR} = (\boldsymbol{R}_{MR})_{\text{diag}} + (\boldsymbol{R}_{MR} - (\boldsymbol{R}_{MR})_{\text{diag}}) $, with the corresponding signal model (\ref{eq.yR3}) re-written as
\begin{equation}\label{eq.yR4}
 \begin{split}
&\tilde{\boldsymbol{Y}}_R  = \underbrace{\left(\boldsymbol{R}_{BR}^{'} \boldsymbol{S}_B  +  (\boldsymbol{R}_{MR})_{\text{diag}} \boldsymbol{S}_M \right)}_{\text{signal to be decoded at the RS}} \\
&+ \underbrace{\left( (\boldsymbol{R}_{BR} -\boldsymbol{R}_{BR}^{'}) \boldsymbol{S}_B  +  (\boldsymbol{R}_{MR} - (\boldsymbol{R}_{MR})_{\text{diag}} ) \boldsymbol{S}_M \right)}_{\text{interference to be cancelled at the RS} }+ \tilde{\boldsymbol{\Psi}}_R,
\end{split}
\end{equation}
where $\boldsymbol{R}_{BR}^{'}$ is an upper-triangular matrix to be determined shortly. The first term in the right hand side (RHS) of (\ref{eq.yR4}), $\tilde{\boldsymbol{S}} \triangleq \left(\boldsymbol{R}_{BR}^{'} \boldsymbol{S}_B  +  (\boldsymbol{R}_{MR})_{\text{diag}} \boldsymbol{S}_M \right) \in {\mathbb C}^{ K \times T}$,  represents the signal to be decoded at the RS.  The second term in the RHS of (\ref{eq.yR4}), $\tilde{\boldsymbol{W}} \triangleq (\boldsymbol{R}_{BR} -\boldsymbol{R}_{BR}^{'}) \boldsymbol{S}_B  +  (\boldsymbol{R}_{MR} - (\boldsymbol{R}_{MR})_{\text{diag}} ) \boldsymbol{S}_M \in {\mathbb C} ^ {K \times T}$, denotes the residual inter-stream interference. We need to properly choose $\boldsymbol{R}_{BR}^{'}$ such that this interference term can be successively cancelled at the RS with decoding ordered from the $K$-th spatial stream to the first stream.  First, to ensure that the second term $\tilde{\boldsymbol{W}}$ in (\ref{eq.yR4}) only contains inter-stream interference, $(\boldsymbol{R}_{BR} -\boldsymbol{R}_{BR}^{'})$ is required to be strictly upper-triangular, i.e., the diagonal of  $\boldsymbol{R}_{BR}^{'}$ is chosen the same as that of $\boldsymbol{R}_{BR}$.  Let $\tilde{\boldsymbol{s}}^{(k)} \in {\mathbb C}^{ 1 \times T}$ and  $\tilde{\boldsymbol{w}}^{(k)} \in {\mathbb C}^{ 1 \times T}$ be the $k$-th row of $\tilde{\boldsymbol{S}}$ and $\tilde{\boldsymbol{W}}$, respectively. In decoding the $k$-th network-coded message $\tilde{\boldsymbol{s}}^{(k)}$, the relay already knows the $(k+1)$-th to $K$-th network-coded messages $\tilde{\boldsymbol{s}}^{(k+1)}, \ldots,\tilde{\boldsymbol{s}}^{(K)}$, since the decoding is ordered from the $K$-th spatial stream to the first stream. If the $k$-th inter-stream interference term $\tilde{\boldsymbol{w}}^{(k)}$  can be expressed as a linear combination of  the $(k+1)$-th to $K$-th network-coded messages $\tilde{\boldsymbol{s}}^{(k+1)}, \ldots,\tilde{\boldsymbol{s}}^{(K)}$, then the relay is able to completely remove $\tilde{\boldsymbol{w}}^{(k)}$ from the received signal. This is  equivalent to say  that there exists a strictly upper-triangular matrix $\boldsymbol{U}_R \in {\mathbb C} ^ {K \times K} $:
 \begin{equation}\label{eq.URD}
 \boldsymbol{U}_R =\left(
                     \begin{array}{ccccc}
                       0 & u_R(1,2) & u_R(1,3)&... & u_R(1,K) \\
                       0 & 0 & u_R(2,3)&... & u_R(2,K) \\
                       ... & ...&... & ... & ... \\
                       0 & ...&... & 0 & u_R(K-1,K) \\
                       0 & 0 & 0 & ... & 0 \\
                     \end{array}
                   \right) ,
\end{equation}
which satisfies
 \begin{equation}\label{eq.UR}
 \begin{split}
 & \boldsymbol{U}_R \tilde{\boldsymbol{S}}= \tilde{\boldsymbol{W}}
 \end{split}
\end{equation}
for arbitrary  $\boldsymbol{S}_B$ and $\boldsymbol{S}_M$. The above condition can be further written as
 \begin{equation}\label{eq.RBR1}
 \boldsymbol{U}_R \boldsymbol{R}_{BR}^{'}  =  \boldsymbol{R}_{BR} -\boldsymbol{R}_{BR}^{'}
 \end{equation}
 \begin{equation}\label{eq.UR2}
 \boldsymbol{U}_R  (\boldsymbol{R}_{MR})_{\text{diag}} =  \boldsymbol{R}_{MR} - (\boldsymbol{R}_{MR})_{\text{diag}}.
\end{equation}
From (\ref{eq.RBR1}), $\boldsymbol{R}_{BR}^{'}$ can be expressed as
 \begin{equation}\label{eq.RBR2}
\boldsymbol{R}_{BR}^{'} = ( \boldsymbol{I}_K + \boldsymbol{U}_R)^{-1} \boldsymbol{R}_{BR}.
 \end{equation}
From (\ref{eq.UR2}), we have
 \begin{equation}\label{eq.UR3}
 \boldsymbol{U}_R   =  (\boldsymbol{R}_{MR} - (\boldsymbol{R}_{MR})_{\text{diag}}) (\boldsymbol{R}_{MR})_{\text{diag}}^{-1}.
\end{equation}
Substituting (\ref{eq.UR3}) into (\ref{eq.RBR2}), we further obtain
 \begin{equation}\label{eq.UR4}
 \begin{split}
 \boldsymbol{R}_{BR}^{'}
 &= ( \boldsymbol{I}_K + (\boldsymbol{R}_{MR} - (\boldsymbol{R}_{MR})_{\text{diag}}) (\boldsymbol{R}_{MR})_{\text{diag}}^{-1})^{-1} \boldsymbol{R}_{BR} \\
 &=  (\boldsymbol{R}_{MR})_{\text{diag}} \boldsymbol{R}_{MR}^{-1} \boldsymbol{R}_{BR}.
 \end{split}
\end{equation}

With such choices of $ \boldsymbol{U}_R$ and  $\boldsymbol{R}_{BR}^{'}$, the relay is then able to remove the inter-stream interference $\tilde{\boldsymbol{W}}$ successively  with decoding ordered from the $K$-th spatial stream $\tilde{\boldsymbol{s}}^{(K)}$ to the first stream $\tilde{\boldsymbol{s}}^{(1)}$.

To see it, let $\tilde{\boldsymbol{y}}_{R}^{(k)}$ and  $\tilde{\boldsymbol{\psi}}_{R}^{(k)}$ be the $k$-th row of $\tilde{\boldsymbol{Y}}_{R}$ and $\tilde{\boldsymbol{\Psi}}_{R}$, respectively. Then, from (\ref{eq.yR4}), the $k$-th subchannel can be expressed as
\begin{equation}\label{eq.yRksub}
\tilde{\boldsymbol{y}}_{R}^{(k)} = \tilde{\boldsymbol{s}}^{(k)}  + \tilde{\boldsymbol{w}}^{(k)}  + \tilde{\boldsymbol{\psi}}_{R}^{(k)}.
\end{equation}

From (\ref{eq.URD}) and (\ref{eq.UR}), the interference term $\tilde{\boldsymbol{w}}^{(k)}$ can be expressed as
\begin{equation}
\tilde{\boldsymbol{w}}^{(k)} = \sum_{n=k+1}^K u_R(k,n) \tilde{\boldsymbol{s}}^{(n)}  ,
\end{equation}
which implies that the interference term $\tilde{\boldsymbol{w}}^{(k)}$ is a linear combination of the signals $\{ \tilde{\boldsymbol{s}}^{(k+1)}, \ldots,  \tilde{\boldsymbol{s}}^{(K)} \}$.  Note that the decoding is ordered from the $K$-th stream to the first stream. When decoding the $k$-th spatial stream $ \tilde{\boldsymbol{s}}^{(k)}$,  $\{ \tilde{\boldsymbol{s}}^{(k+1)}, \ldots,  \tilde{\boldsymbol{s}}^{(K)}\}$ are already decoded and are known to the relay. As a result, the residue ISI $\tilde{\boldsymbol{w}}^{(k)}$ can be completely removed from the received signal $\tilde{\boldsymbol{y}}^{(k)}$:
\begin{equation}
\tilde{\boldsymbol{z}}_R^{(k)} = \tilde{\boldsymbol{y}}_R^{(k)} -  \tilde{\boldsymbol{w}}^{(k)} =  \tilde{\boldsymbol{s}}^{(k)}  + \tilde{\boldsymbol{\psi}}_R^{(k)}.
\end{equation}


 After successive interference cancellation, the $k$-th subchannel scaled by the factor $ \frac{1}{r_{BR}(k,k)}$, can be expressed as
\begin{equation}\label{eq.yRkn}
\begin{split}
\boldsymbol{y}_{R,k} &=  \frac{\tilde{\boldsymbol{z}}_R^{(k)}}{r_{BR}(k,k)}  = \frac{\tilde{\boldsymbol{s}}^{(k)}  + \tilde{\boldsymbol{\psi}}_{R}^{(k)}}{r_{BR}(k,k)} \\
 &=  \boldsymbol{s}_{B,k}  +  \alpha_k \boldsymbol{s}_{M,k}
 +  \boldsymbol{v}_{k} + \boldsymbol{\psi}_{R,k},
 \end{split}
\end{equation}
where
\begin{subequations}\label{eq.uva}
\begin{align}
&\alpha_k = \frac{ r_{MR}(k,k)}{r_{BR}(k,k)},  \label{eq.uva.c}\\
&\boldsymbol{v}_{k} = \sum_{j=k+1}^K \frac{r_{BR}^{'}(k,j)}{r_{BR}(k,k)} \boldsymbol{s}_{B,j}, \label{eq.uva.b}\\
&\boldsymbol{\psi}_{R,k} = \tilde{\boldsymbol{\psi}}_{R}^{(k)}/r_{BR}(k,k). \label{eq.uva.d}
\end{align}
\end{subequations}

 Note that the decoded signal of the $k$-th stream is $\boldsymbol{s}_{B,k}  +  \alpha_k \boldsymbol{s}_{M,k}
 + \boldsymbol{v}_{k} $.  Here, $\boldsymbol{v}_{k}$ in (\ref{eq.uva.b}) is a weighted sum of the last $(K-k)$ rows of $\boldsymbol{S}_B$. Let the encoding order of the BS be also from the $K$-th spatial stream to the first stream. Then, when encoding the $k$-th spatial stream at the BS, $\boldsymbol{v}_{k}$ is known to BS; thus it can be pre-cancelled using lattice precoding at the BS, as will be elaborated in the sequel.

\subsection{Phase 1: Encoding at the BS and MSs}
In the following, we describe the encoding and decoding operations involved in the proposed scheme. We start with the encoding operations at the BS and MSs.

Let ${\cal{W}}_{M,k} = \{1,2,...,2^{TR_{M,k}} \}$ be the message set for the data stream at the $k$-th MS, and $w_{M,k} \in {\cal{W}}_{M,k}$ be the corresponding message. The $(1 \times T)$-dimensional coded vector for the spatial data stream at the $k$-th MS is denoted as $\boldsymbol{s}_{M,k}=f_{M,k}(w_{M,k})$, where $f_{M,k}(\cdot)$ is the encoding function to be specified in the following. Similarly, the message of the $k$-th spatial stream transmitted to the $k$-th MS at the BS is denoted by $w_{B,k} \in {\cal{W}}_{B,k}$, where ${\cal{W}}_{B,k} = \{1,2,...,2^{TR_{B,k}} \}$ is the message set. The corresponding $(1 \times T)$-dimensional encoded vector is denoted by $\boldsymbol{s}_{B,k} = f_{B,k}(w_{B,k})$.

Nested lattice coding is applied to each message pair $(w_{B,k},w_{M,k})$. An $n$-dimensional lattice $\Lambda$ is a subgroup of $\mathbb{R}^{n}$ under normal vector addition. A lattice $\Lambda$ is {\it nested} in the lattice $\Lambda_1$ if $\Lambda \subseteq \Lambda_1$. The main idea of nested lattice codes is to use the coarse lattice $\Lambda$ as a shaping region and the lattice points from the fine lattice $\Lambda_1$ within the Voronoi region of the coarse lattice $\Lambda$ as the codewords. The details of nested lattice codes can be found in \cite{Zam02,Erez04,Erez05,Nam10}.

The encoding functions $\{f_{B,k}(\cdot)\}_{k=1}^K$ and $\{f_{M,k}(\cdot)\}_{k=1}^K$ are described as follows. The encoding of the BS follows the order from the $K$-th stream to the first stream sequentially. Without loss of generality, we assume $R_{B,k} \le R_{M,k}$. For the $k$-th subchannel in (\ref{eq.yRkn}), we construct a nested lattice chain $\Lambda_{B,k}$, $\Lambda_{M,k}$ and $\Lambda_{C,k}$ satisfying $\Lambda_{M,k} \subseteq \Lambda_{B,k} \subseteq \Lambda_{C,k}$. Here, $\Lambda_{B,k}$ and $\Lambda_{M,k}$ are simultaneously Rogers-good and Poltyrev-good while $\Lambda_{C,k}$ is Poltyrev-good \cite{Nam10}. Let $\boldsymbol{c}_{{\cal{X}},k} \in \mathcal{C}_{{\cal{X}},k} $ denote the codeword mapped from the message $w_{{\cal{X}},k}$, ${\cal{X}} \in \{B,M\}$, where $\mathcal{C}_{{\cal{X}},k}$ is the nested lattice code defined by $\Lambda_{{\cal{X}},k}$ and $\Lambda_{C,k}$.  The coding rate of the nested lattice code $\mathcal{C}_{{\cal{X}},k}$  can approach any ${\xi}_{\cal{X}}>0$ as $T \to \infty$ \cite{Nam10}, i.e.
\begin{equation}
R_{{\cal{X}},k} = \frac{1}{T} \log \left(\frac{\text{Vol}(\Lambda_{{\cal{X}},k})}{\text{Vol}(\Lambda_{C,k})}\right) = {\xi}_{\cal{X}} + o_{{T}}(1),
\end{equation}
where $\text{Vol}(\Lambda)$ is the volume of the Voronoi region of a
lattice $\Lambda$, and  $o_T(1) \to 0$ as $T \to \infty$.

For the $k$-th subchannel in (\ref{eq.yRkn}), $\Lambda_{B,k}$ and $\Lambda_{M,k}$ are  chosen to meet
\begin{equation}
 \left( \frac{\text{Vol}(\Lambda_{M,k})}{\text{Vol}(\Lambda_{B,k})} \right) ^{\frac{1}{T}} =  \left(  \frac{|r_{MR}(k,k)|^2 P_{M,k}} {|r_{BR}(k,k)|^2 P_{B,k}} \right) ^{\frac{1}{2}}+ o_{{T}}(1),
\end{equation}
where $P_{B,k}$ denotes the average power of the $k$-th spatial stream at the BS. Then, the relation between $R_{B,k}$ and $R_{M,k}$ can be written as \cite{Nam10}
\begin{equation}\label{eq.ratebm}
R_{B,k} = R_{M,k} +  \frac{1}{2} \log \left( \frac{|r_{BR}(k,k)|^2 P_{B,k}} {|r_{MR}(k,k)|^2 P_{M,k}}\right) + o_{T}(1).
\end{equation}

The encoding at the BS is as follows. Let $\boldsymbol{d}_{{\cal{X}},k}$ be a random dithering vector that is uniformly distributed over the Voronoi region of $\Lambda_{{\cal{X}},k}$, ${\cal{X}} \in \{B,M\}$\footnote{It is always assumed that the dithering signals, such as $\boldsymbol{d}_{{\cal{X}},k}$, are globally known to all the nodes in the network.}.  With random dithering,  the $k$-th transmit signal $\boldsymbol{s}_{B,k}$ at the BS is constructed as
\begin{equation}\label{eq.lattice.bkn}
\boldsymbol{s}_{B,k} = \left(\boldsymbol{c}_{B,k} - \boldsymbol{v}_{k} - \boldsymbol{d}_{B,k} \right) ~\text{mod}~ \Lambda_{B,k},
\end{equation}
where the inter-stream interference $\boldsymbol{v}_{k}$ is  {\it a-priori} known when encoding the $k$-th stream at the BS (cf. (\ref{eq.uva.b})).

The signal $\boldsymbol{s}_{M,k}$ at the $k$-th MS is encoded as
\begin{equation}\label{eq.lattice.mkn}
\boldsymbol{s}_{M,k} = \frac{1}{\alpha_k} \left( \left(\boldsymbol{c}_{M,k} - \boldsymbol{d}_{M,k} \right) ~\text{mod}~ \Lambda_{M,k} \right) ,
\end{equation}
where $\boldsymbol{d}_{M,k}$ is a random dithering vector uniformly distributed over the Voronoi region of $\Lambda_{M,k}$.

It is worth mentioning that the authors in \cite{YangHJ12} presumed that random dithering cannot be used in the first phase of the MIMO cTWRC due to the non-cooperation among MSs. Interestingly, we will show that through a careful precoder design, random dithering can be in fact employed to improve performance.

\subsection{Relay's Operation: Lattice Decoding}
We now consider the operations at the relay. The relay's decoding order is from the $K$-th stream to the first stream. For the $k$-th subchannel in (\ref{eq.yRkn}), the relay intends to decode the combinations $\boldsymbol{s}_{B,k} + \alpha_k \boldsymbol{s}_{M,k} + \boldsymbol{v}_{k}$. Recall that $\tilde{\boldsymbol{w}}^{(k)}$  in (\ref{eq.yRksub}) is a weighted sum of the network-coded signals already decoded; hence it can be cancelled  from the received signal before Lattice decoding. Together with the knowledge of the dithering vectors $\boldsymbol{d}_{B,k}$ and  $\boldsymbol{d}_{M,k}$, the relay constructs
\begin{subequations}\label{eq.yRkn2}
\begin{align}
 \boldsymbol{w}_{R,k} & = \boldsymbol{y}_{R,k}  + \boldsymbol{d}_{B,k} +  \boldsymbol{d}_{M,k} \\
 &= \boldsymbol{s}_{B,k} \! + \! \boldsymbol{d}_{B,k}  \! +  \! \boldsymbol{v}_{k}\!\!+\!\!  \alpha_k \boldsymbol{s}_{M,k}\!\! +\!\!\boldsymbol{d}_{M,k}
 \!\!+\!\! \boldsymbol{\psi}_{R,k}\\
 &= \tilde{\boldsymbol{c}}_{B,k} + \tilde{\boldsymbol{c}}_{M,k} +  \boldsymbol{\psi}_{R,k},
 \end{align}
\end{subequations}
where $\tilde{\boldsymbol{c}}_{B,k} = \boldsymbol{s}_{B,k} + \boldsymbol{d}_{B,k} +  \boldsymbol{v}_{k}$, and $ \tilde{\boldsymbol{c}}_{M,k} = \alpha_k \boldsymbol{s}_{M,k} +\boldsymbol{d}_{M,k}$. From (\ref{eq.lattice.bkn}), we see that $\tilde{\boldsymbol{c}}_{B,k} = (\boldsymbol{c}_{B,k} - \boldsymbol{v}_{k} - \boldsymbol{d}_{B,k}) ~\text{mod}~ \Lambda_{B,k} + \boldsymbol{d}_{B,k} +  \boldsymbol{v}_{k}$ is a lattice point in $ \Lambda_{C,k}$. Similarly, $\tilde{\boldsymbol{c}}_{M,k} = (\boldsymbol{c}_{M,k} - \boldsymbol{d}_{M,k}) ~\text{mod}~ \Lambda_{M,k} +\boldsymbol{d}_{M,k} $ is a lattice point in $\Lambda_{C,k}$. Hence, $(\tilde{\boldsymbol{c}}_{B,k} + \tilde{\boldsymbol{c}}_{M,k}) \in \Lambda_{C,k} $ is decodable using lattice decoding with a vanishing error probability, provided that \cite{Yuan13}
\begin{equation}\label{eq.ulratebr}
R_{B,k} \le R_{B \to R,k}(P_{B,k}) \triangleq \left[ \frac{1}{2} \log\left( \frac{|r_{BR}(k,k)|^2 P_{B,k} }{\sigma^2}\right) \right]^+.
\end{equation}
From (\ref{eq.ratebm}), we also obtain
\begin{equation}\label{eq.ulratemr}
R_{M,k} \le R_{M \to R,k}(P_{M,k}) \triangleq \left[ \frac{1}{2} \log\left( \frac{|r_{MR}(k,k)|^2 P_{M,k} }{\sigma^2}\right) \right]^+.
\end{equation}

Once the lattice point $\tilde{\boldsymbol{c}}_{B,k} + \tilde{\boldsymbol{c}}_{M,k}$ is decoded (with a vanishing error probability as $T \to \infty $), the relay can reconstruct  $\boldsymbol{s}_{B,k} + \alpha_k \boldsymbol{s}_{M,k}  +  \boldsymbol{v}_{k} = \tilde{\boldsymbol{c}}_{B,k} + \tilde{\boldsymbol{c}}_{M,k} - \boldsymbol{d}_{B,k} - \boldsymbol{d}_{M,k}$, which is used as the known signal to cancel the residual inter-stream interference in subsequent decoding; see (\ref{eq.yRksub}) and the discussions therein.

\subsection{Relay's Operation: Re-encoding}

After lattice decoding, the relay calculates
\begin{equation}\label{eq.sr}
\boldsymbol{s}_{R,k} = \left( \tilde{\boldsymbol{c}}_{B,k} +   \tilde{\boldsymbol{c}}_{M,k} \right)  ~\text{mod}~ \Lambda_{B,k},
 \end{equation}
for $k=1,\ldots,K$. Then, one-side DPC encoding is applied to $\boldsymbol{s}_{R,1},\ldots,\boldsymbol{s}_{R,K}$,  with an order $\mu_1, \ldots, \mu_K$, so that each MS receives an interference-free signal. Note that the decoding order $\boldsymbol{\mu} = [\mu_1, \ldots, \mu_K]^T$ is specified by $\boldsymbol{\Phi}$ as $\boldsymbol{\phi}(k,\mu_k)=1$, and $\boldsymbol{\phi}(k,j)=0, \forall j \neq \mu_k, k=1, \ldots,K$.  From (\ref{eq.ym2}), the received signal at the $k$-th MS can be expressed as
\begin{equation}\label{eq.ymk}
\boldsymbol{y}_{M,k} = l_{RM}(q_k,q_k)\boldsymbol{x}_{R,DPC,q_k} + \boldsymbol{t}_{M,k} + \tilde{\boldsymbol{\psi}}_{M}^{(q_k)},
\end{equation}
where $\mu_{q_k} = k$, $\tilde{\boldsymbol{\psi}}_{M}^{(q_k)}$ denotes the $q_k$-th row of $\tilde{\boldsymbol{\Psi}}_{M}$,  $\boldsymbol{t}_{M,k} = \sum_{n=1}^{q_k-1} l_{RM}(q_k,n)\boldsymbol{x}_{R,DPC,n}$ and $\boldsymbol{x}_{R,DPC,n}$ denotes the $n$-th DPC encoded signal. The interference $\boldsymbol{t}_{M,k}$ can be pre-cancelled at the RS using dirty paper precoding as
$$
\boldsymbol{x}_{R,DPC,q_k} = \left( \boldsymbol{s}_{R,k} - \beta_{M,k} \frac{\boldsymbol{t}_{M,k}}{l_{RM}(q_k,q_k)}  - \boldsymbol{d}_{R,k} \right) ~\text{mod}~ \Lambda_{B,k},
$$
where $\boldsymbol{d}_{R,k}$ is a random dither vector that is known by the BS and the MSs, and $\beta_{M,k} = |l_{RM}(q_k,q_k)|^2 P_{R,k} $ $/ (|l_{RM}(q_k,q_k)|^2 P_{R,k} + \sigma^2)$ is the minimum mean square error coefficient for decoding at the MS \cite{Erez04}, with $P_{R,k}$ being the transmit power of the $k$-th stream of the RS satisfying $\sum_{k=1}^{K}  P_{R,k} \leq P_R$. The DPC encoded signal $\boldsymbol{X}_{R,DPC} = [\boldsymbol{x}_{R,DPC,1}^T, \ldots,\boldsymbol{x}_{R,DPC,K}^T]^T$ is then linearly precoded as (\ref{eq.xr}) and broadcast to the BS and MSs in the second phase.

\subsection{Phase 2: MS Decoding}
The decoding operation at each MS is as follows: For $k=1,\ldots,K$, the $k$-th MS first decode $\boldsymbol{s}_{R,k}$ from the received signal $\boldsymbol{y}_{M,k}$; then it recovers $ \boldsymbol{c}_{B,k}$ from $\hat{\boldsymbol{s}}_{R,k}$ (i.e., the decoded $\boldsymbol{s}_{R,k}$) with the help of the knowledge of the self-message $\boldsymbol{c}_{M,k}$ and the dither signal $\boldsymbol{d}_{M,k}$. For the first step, it has been shown in \cite{YangHJ12} that the probability
$\hat{\boldsymbol{s}}_{R,k} \neq \boldsymbol{s}_{R,k}$  vanishes as $T \to \infty$ provided that
\begin{equation}\label{eq.dlratems}
R_{B,k} \le R_{R \to M,k}(P_{R,k}) \triangleq  \frac{1}{2} \log\left(1 +  \frac{|l_{RM}(q_k,q_k)|^2 P_{R,k} }{\sigma^2}\right).
\end{equation}
Here we focus on the second step, i.e., to recover  $ \boldsymbol{c}_{B,k}$ from $\hat{\boldsymbol{s}}_{R,k}$. Note that  $\boldsymbol{s}_{R,k}$ in (\ref{eq.sr}) can be written as
\begin{equation}\label{eq.sr2}
\begin{split}
&\boldsymbol{s}_{R,k} = \left( \tilde{\boldsymbol{c}}_{B,k} +   \tilde{\boldsymbol{c}}_{M,k} \right)  ~\text{mod}~ \Lambda_{B,k} \\
&= \left( \boldsymbol{s}_{B,k} \!+\! \boldsymbol{d}_{B,k} \!+\!  \boldsymbol{v}_{k} \!+\! \alpha_k \boldsymbol{s}_{M,k} \!+\! \boldsymbol{d}_{M,k} \right)  ~\text{mod}~ \Lambda_{B,k} \\
&= \left( \boldsymbol{c}_{B,k} \!+\!  \left( \boldsymbol{c}_{M,k}\!-\! \boldsymbol{d}_{M,k} \right) ~\text{mod}~ \Lambda_{M,k} \!+\! \boldsymbol{d}_{M,k} \right)  ~\text{mod}~ \Lambda_{B,k}.
\end{split}
\end{equation}

As $\boldsymbol{c}_{M,k}$ and $\boldsymbol{d}_{M,k}$ are known, the $k$-th MS obtains $\boldsymbol{c}_{B,k}$ as
\begin{equation}\label{eq.mdec}
\begin{split}
\hat{\boldsymbol{c}}_{B,k} &= \left(  \hat{\boldsymbol{s}}_{R,k} \!-\!  \left(\boldsymbol{c}_{M,k} \!-\! \boldsymbol{d}_{M,k} \right) ~\text{mod}  \Lambda_{M,k} \!-\! \boldsymbol{d}_{M,k} \right) ~\text{mod}  \Lambda_{B,k}  \\
&= \left(  \hat{\boldsymbol{s}}_{R,k} - \boldsymbol{s}_{R,k}  + \boldsymbol{c}_{B,k} \right) ~\text{mod}  \Lambda_{B,k}  \\
&= \boldsymbol{c}_{B,k},
\end{split}
\end{equation}
where the second equality follows from
$$
\left( \boldsymbol{x} ~\text{mod} ~ \Lambda_{M,k} \right) ~\text{mod} ~ \Lambda_{B,k} =  \boldsymbol{x} ~\text{mod} ~ \Lambda_{B,k}
$$
 as $\Lambda_{M,k} \subseteq \Lambda_{B,k}$ , and the last equality holds provided $\hat{\boldsymbol{s}}_{R,k} = \boldsymbol{s}_{R,k}$.

\subsection{Phase 2: BS Decoding}

The BS first decodes $\boldsymbol{s}_{R,k},k=1,\ldots,K$, from the received signal $\tilde{\boldsymbol{Y}}_{B}$ in (\ref{eq.yb2}). Note that inter-stream interference still exists at the BS since DPC encoding is only applied to the RS-MS link. However, if the decoding order at the BS is the same as the encoding order at the RS, it was shown in \cite{YangHJ12} that this interference can be successively cancelled at the BS since $\boldsymbol{L}_{RB}$ is lower-triangular. The corresponding probability of $\hat{\boldsymbol{s}}^{'}_{R,k} \neq \boldsymbol{s}_{R,k}$  vanishes as $T \to \infty$, provided that
\begin{equation}\label{eq.dlratebs}
R_{M,k} \le R_{R \to B,k}(P_{R,k}) \triangleq \frac{1}{2} \log\left(1 +  \frac{|l_{RB}(q_k,q_k)|^2 P_{R,k} }{\sigma^2}\right).
\end{equation}

With the knowledge of the self-message $\boldsymbol{c}_{B,k}$, the BS then recovers $\boldsymbol{c}_{M,k}$ from $\hat{\boldsymbol{s}}_{R,k}^{'}$ by calculating
\begin{equation}\label{eq.bdec1}
\hat{\boldsymbol{c}}_{M,k} = \left(  \hat{\boldsymbol{s}}_{R,k}^{'}  - \boldsymbol{c}_{B,k} \right) ~\text{mod} ~ \Lambda_{M,k}.
\end{equation}

To see this, we obtain from (\ref{eq.sr}) that
\begin{equation}\label{eq.bdec2}
\begin{split}
\hat{\boldsymbol{c}}_{M,k} &\!\!=\!\! \left( \hat{\boldsymbol{s}}_{R,k}^{'} \!\!-\!\!  \boldsymbol{s}_{R,k} \!\!+\!\! (\boldsymbol{c}_{M,k} \!\!-\!\! \boldsymbol{d}_{M,k}) \text{mod} \Lambda_{M,k} +  \boldsymbol{d}_{M,k}  \right) \text{mod}  \Lambda_{M,k}\\
&= \left(\hat{\boldsymbol{s}}_{R,k}^{'} -  \boldsymbol{s}_{R,k} + \boldsymbol{c}_{M,k} \right) ~\text{mod}~ \Lambda_{M,k} \\
&= \boldsymbol{c}_{M,k},
\end{split}
\end{equation}
where the last step holds provided $\hat{\boldsymbol{s}}_{R,k}^{'} = \boldsymbol{s}_{R,k}$.

\subsection{Achievable Rates of the Overall Scheme}
Combining the discussions in Subsections $C\text{-}G$, we have the following theorem for the proposed two-way relaying scheme.
\theorem As $T \to + \infty$, a rate tuple of $(R_{B,1},\ldots, R_{B,K}$, $R_{M,1},\ldots,R_{M,K})$ of the MIMO cTWRC is achievable if
\begin{equation}\label{eq.ratetuple}
\begin{split}
R_{B,k} &\le \min \left(R_{B \to R,k}(P_{B,k}), R_{R \to M,k}(P_{R,k}) \right), \\
R_{M,k} &\le \min \left(R_{M \to R,k}(P_{M,k}), R_{R \to B,k}(P_{R,k}) \right),
\end{split}
\end{equation}
$k=1,\ldots,K$, where $R_{B \to R,k}(P_{B,k})$, $R_{R \to M,k}(P_{R,k})$, $R_{M \to R,k}(P_{M,k})$ and $R_{R \to B,k}(P_{R,k})$ are given in (\ref{eq.ulratebr}), (\ref{eq.dlratems}), (\ref{eq.ulratemr}) and (\ref{eq.dlratebs}), respectively.

\section{Analysis of the Sum-Rate Performance}

In this section, we analyze the sum-rate performance of the proposed two-way relaying scheme in the high SNR regime.  We first consider the cut-set bound in (\ref{eq.cutset1}). It is known that, in the high SNR regime,  equal power allocation at the BS and the RS is optimal, i.e., $\boldsymbol{Q}_B^{opt} = \frac{P_B}{K}\boldsymbol{I}_K$, and $\boldsymbol{Q}_R^{opt} = \frac{P_R}{K}\boldsymbol{I}_K$. Also, as the MSs can not cooperate, the optimal $\boldsymbol{Q}_M$ is given by $\boldsymbol{Q}_M^{opt} = \text{diag}(P_{M,1}, \ldots, P_{M,K})$. Then the sum-rate of cut-set bound at high SNR can be written as
\begin{equation}\label{eq.cutset}
\begin{split}
&R_{\text{sum,cs}} = \sum_{k=1}^{K}   \left( R_{B,k} + R_{M,k} \right) \\
&\simeq  \min \left( \sum_{k=1}^{K}  \frac{1}{2} \log \left( \frac{\lambda_{BR,k} P_B} {K \sigma^2} \right), \sum_{k=1}^{K}  \frac{1}{2} \log \left( \frac{\lambda_{RM,k} P_R} {K \sigma^2} \right) \right) \nonumber \\
&+  \min \left( \sum_{k=1}^{K}  \frac{1}{2} \log \left( \frac{\lambda_{MR,k} P_{M,k}} { \sigma^2} \right), \sum_{k=1}^{K}  \frac{1}{2} \log \left( \frac{\lambda_{RB,k} P_R} {K \sigma^2} \right) \right)
\end{split}
\end{equation}
where $\lambda_{{\cal{X}},k}$ denotes the square of the $k$-th singular value of $\boldsymbol{H}_{\cal{X}}$, ${\cal{X}} \in \{ \{BR\},\{MR\},\{RB\},\{RM\} \}$, and $x \simeq y $ means $x - y \to 0 $ as $ \sigma^2 \to 0$.

We now consider the proposed two-way relaying scheme. From (\ref{eq.ratetuple}), the achievable sum-rate of the proposed scheme is given by
\begin{equation}\label{eq.sum-rate}
\begin{split}
R_{\text{sum}} &=  \sum_{k=1}^{K}  [ \min \left(R_{B \to R,k}(P_{B,k}), R_{R \to M,k}(P_{R,k}) \right) \\
&\quad +  \min \left(R_{M \to R,k}(P_{M,k}), R_{R \to B,k}(P_{R,k}) \right) ].
\end{split}
\end{equation}
It is difficult to derive a closed-form expression for the optimal power allocation, even in the high SNR regime. In the following, we assume equal power allocation, i.e., $P_{B,k}=P_B/K,P_{R,k}=P_R/K, \forall k$, which in general gives a lower-bound of achievable sum-rate. We next show that the proposed scheme with equal power allocation can asymptotically achieve the cut-set bound (\ref{eq.cutset}) under certain conditions.

To start with, we consider the situation that the transmission rate of the BS-to-MS link is bottle-necked by the BS-RS link. For the cut-set bound (\ref{eq.cutset}), it is not difficult to see that the achievable sum-rate of the BS-RS link is no greater than the relayed RS-MS link if the transmit power of the BS satisfies
\begin{equation}\label{eq.fpr}
P_B \le \rho_{B}  P_R,
\end{equation}
where $\rho_{B} = \sqrt[K]{ \prod_{k=1}^{K} ({\lambda_{RM,k}}/{\lambda_{BR,k}}})$. On the other hand,  for the proposed scheme with equal power allocation, the transmission rate of the $k$-th spatial stream from the BS to the RS is less than or equal to that from the RS to the $k$-th MS if
\begin{equation}\label{eq.fprk}
P_B \le   \rho_{B,k} P_R,
\end{equation}
where $\rho_{B,k} ={|l_{RM}(q_k,q_k)|^2}/{|r_{BR}(k,k)|^2}$. We will show that if both (\ref{eq.fpr}) and (\ref{eq.fprk}) hold for all the $K$ spatial streams, i.e., $P_B \le \min \left( \rho_{B}, \min_{k} \rho_{B,k}\right) P_R$,  then the proposed scheme achieves the sum-rate cut-set bound of the BS-to-MS link in the high SNR regime. Similarly, the proposed scheme asymptotically achieves the sum-rate capacity if the data transmission of the BS-to-MS link is bottle-necked by the RS-MS link. Furthermore, it can be shown that the cut-set bound of the MS-to-BS link can be achieved in the high SNR regime if the data transmission from the MSs to the BS is bottle-necked by either the MS-RS link or the RS-BS link. Following these lines, we define the conditions C1-C4 as follows:

C1: $P_B \le \min \left( \rho_{B}, \min_{k} \rho_{B,k}\right) P_R $;

C2: $P_B \ge \max \left( \rho_{B}, \max_{k} \rho_{B,k}\right) P_R $;

C3: $ P_R \le \min \left(\rho_{M} { \prod_{k=1}^{K} P_{M,k}^{1/K} }, \min_{k} \rho_{M,k} P_{M,k} \right)$;

C4: $ P_R \ge \max \left(\rho_{M} { \prod_{k=1}^{K} P_{M,k}^{1/K} }, \max_{k} \rho_{M,k} P_{M,k} \right)$, \\
where $\rho_{M} = \sqrt[K]{ \prod_{k=1}^{K} ({\lambda_{MR,k}}/{\lambda_{RB,k}}})$, and
$\rho_{M,k} ={|r_{MR}(k,k)|^2}/{|l_{RB}(q_k,q_k)|^2}$. Then  we can establish that:

\theorem The proposed scheme achieves the cut-set bound (\ref{eq.cutset}) if one of the conditions C1-C2 holds and one of the conditions C3-C4 holds for a given DPC order $\boldsymbol{\Phi}$.

\proof See Appendix A.

Note that the conditions C1-C4 depend on the DPC order at the RS (i.e. different encoding order  $\boldsymbol{\Phi}$ results in different conditions). Theorem 2 states that the proposed scheme is able to asymptotically achieve the sum capacity of the MIMO cTWRC, provided that there exists such a DPC  order that  the data exchanges between the BS and the MSs are simultaneously constrained by either the BS-RS link or the MS-RS link. In contrast, for the precoding and decoding scheme proposed by Yang in \cite{YangHJ12}, it is required that the SNR from the BS to the RS is much higher than the SNR from the RS to the MSs, such that the power inefficiency incurred by the precoding operation at the BS is negligible. Hence, the Yang's scheme in \cite{YangHJ12} can only asymptotically achieve the sum capacity when the data exchanges are simultaneously bottle-necked by the MS-RS link.  For the proposed novel precoding/decoding design, only unitary transforms are involved for the channels, and interference-free PNC can be performed with QR decomposition based SIC at the relay\footnote{It is known that QR based SIC is sufficient to achieve the cut-set bound in the high SNR regime \cite{Cai03,YangHJ12}.}. As a result, the asymmetry between the BS-RS and MS-RS links is no longer a prerequisite to achieve near-capacity performance for the proposed scheme.


\section{Optimal Power Allocation}
We have shown that, under certain conditions, the proposed two-way relaying scheme achieves the sum-capacity of  the MIMO cTWRC with equal power allocation in the high SNR regime. To fully exploit the potential of the proposed scheme, we next investigate the optimal power allocation at the BS and the RS to maximize the weighted sum-rate of the MIMO cTWRC at general SNR (which includes the sum-rate in (\ref{eq.sum-rate}) as a special case where all weights are equal). We will show that this optimization problem can be formulated as a monotonic program and can be solved by a polyblock outer approximation algorithm.

It is clear that, the optimal power allocation for each MS is to transmit with maximum power $P_{M,k}$. Let $\boldsymbol{P}_B := [P_{B,1},\ldots,P_{B,K}]^T$ and $\boldsymbol{P}_R := [P_{R,1},\ldots,P_{R,K}]^T$ be the power allocation profiles at the BS and the RS, respectively, and   $\boldsymbol{P} := [\boldsymbol{P}_B^T,\boldsymbol{P}_R^T]^T$. Consider the following weighted sum-rate maximization problem:
\begin{subequations}\label{eq.maxsumrate}
\begin{align}
& \max_{\boldsymbol{P}} R_{\text{ws}} = \sum_{k=1}^{K}  \left( \xi_{B,k} R_{B,k}  +   \xi_{M,k} R_{M,k} \right) \\
& \text{s. t.}  ~~~  \sum_{k=1}^{K}  P_{B,k} \leq P_B, ~ \sum_{k=1}^{K}  P_{R,k} \leq P_R.
\end{align}
\end{subequations}
where $\xi_{B,k}$ and $\xi_{M,k}$ denote the weights assigned to the $k$-th data stream at the BS and the data stream of the $k$-th MS, respectively.

The problem (\ref{eq.maxsumrate}) is not  convex. Yet, it can be reformulated as a monotonic program, which could be efficiently solved by a polyblock outer approximation method \cite{Tuy00}.  Using $ [\log(x)]^+ = \log(1+(x-1)^+)$,  the achievable rate $R_{B,k}$ in (\ref{eq.ratetuple}) can be expressed as
\begin{equation}\label{eq.rbk}
R_{B,k} = \frac{1}{2} \log (1 + SNR_{BM,k}(\boldsymbol{P})),
\end{equation}
where
\begin{equation}\label{eq.snrbm}
\begin{split}
SNR_{BM,k}(\boldsymbol{P}) =& \min \left(\left(\frac{|r_{BR}(k,k)|^2 P_{B,k} }{\sigma^2} -1 \right)^{+} , \right. \\
  & \left. \quad \quad \frac{|l_{RM}(q_k,q_k)|^2 P_{R,k} }{\sigma^2} \right) .
\end{split}
\end{equation}

Similarly, $R_{M,k}$ in (\ref{eq.ratetuple}) can be expressed as
\begin{equation}\label{eq.rmk}
R_{M,k} = \frac{1}{2} \log (1 + SNR_{MB,k}(\boldsymbol{P})),
\end{equation}
where
\begin{equation}\label{eq.snrmb}
\begin{split}
SNR_{MB,k}(\boldsymbol{P}) =& \min \left(\left(\frac{|r_{MR}(k,k)|^2 P_{M,k} }{\sigma^2} -1 \right)^{+} , \right. \\
& \left. \quad \quad \frac{|l_{RB}(q_k,q_k)|^2 P_{R,k} }{\sigma^2} \right).
\end{split}
\end{equation}

Then, the weighted sum-rate $R_{\text{ws}}$ in (\ref{eq.maxsumrate}) can be expressed as
\begin{equation}\label{eq.maxsumrate2}
 R_{\text{ws}} = \sum_{i=1}^{2K} \frac{\xi_i}{2} \log(1+ \text{SNR}_i(\boldsymbol{P})).
\end{equation}
where $\xi_i$ is the $i$-th element of the weight vector $\boldsymbol{\xi} = [\xi_{B,1},\ldots,\xi_{B,K},\xi_{M,1},\ldots,\xi_{M,K}]^T$, $\text{SNR}_{i}(\boldsymbol{P}) = \text{SNR}_{BM,i}(\boldsymbol{P})$, and $\text{SNR}_{K + i}(\boldsymbol{P}) = \text{SNR}_{MB,i}(\boldsymbol{P})$, for $i=1,\ldots,K$.

Define the set ${\cal S} := \{ \boldsymbol{P} | \sum_{k=1}^{K}  P_{B,k} \leq P_B, ~ \sum_{k=1}^{K}  P_{R,k} \leq P_R, ~ k=1,\ldots,K\}$. Introducing an auxiliary vector $\boldsymbol{z} = [z_1,\ldots,z_{2K}]^T$,
we can rewrite (\ref{eq.maxsumrate}) as
\begin{equation}\label{eq.mp1}
\max_{\boldsymbol{z} \in {\cal Z}} \Gamma(\boldsymbol{z}) := \sum_{i=1}^{2K} \frac{\xi_i}{2}\log (z_i).
\end{equation}
where the feasible set $ {\cal Z} :=\{\boldsymbol{z}|1 \leq z_i \leq 1 +
\text{SNR}_i(\boldsymbol{P}),
i=1, \ldots, 2K, \forall \boldsymbol{P} \in {\cal S} \} $. Let
\begin{equation}\label{GG}
{\cal G}:= \{\boldsymbol{z}\;|\; 0 \leq z_i \leq 1 +
\text{SNR}_i(\boldsymbol{P}),
i=1, \ldots, 2K, \forall \boldsymbol{P} \in {\cal S} \}.
\end{equation}
It can be shown that ${\cal G}$ is a compact normal set with nonempty interior \cite{Tuy00}. Further ${\cal H}:=\{\boldsymbol{z}\;|\; z_i \geq 1, \forall
i\}$ is a reverse normal set. Then,
(\ref{eq.mp1}) becomes a standard MP
\cite{Tuy00} as
\begin{equation}\label{eq.mp2}
\max_{\boldsymbol{z}} \;\Gamma(\boldsymbol{z}) \quad \text{s. t.}  ~~ \boldsymbol{z} \in  {\cal G} \cap {\cal H}.
\end{equation}

For the MP in (\ref{eq.mp2}),  we can use a polyblock outer approximation method to find its global optimal solution \cite{Tuy00}. This method has been used for power control \cite{Qian09}, multi-cell coordinated beamforming \cite{Uts12}, etc. The main idea of the iterative polyblock outer approximation algorithm is to construct a series of outer polyblocks ${\cal P}_n$ to approximate ${\cal G} \cap {\cal H}$. Given any finite set ${\cal T}_n = \{\boldsymbol{v}_i | i=1,\ldots,I \} $, the union of all the sets $\{\boldsymbol{x}|0 \leq \boldsymbol{x} \leq \boldsymbol{v}_i \}$ is a  {\it polyblock} with  {\it vertex} set ${\cal T}_n$. A polyblock ${\cal {P}}_n $ is an {\it outer polyblock} of ${\cal {S}} $ if $ {\cal {S}} \subseteq {\cal {P}}_n$ \cite{Tuy00}. Usually, the algorithm
starts from a one-vertex outer polyblock ${\cal P}_0$ of ${\cal G} \cap {\cal H}$, and a smaller new outer polyblock is constructed in each iteration. A key step in the construction of the new outer polyblock ${\cal {P}}_{n+1} $ from ${\cal {P}}_n $ is to find the following projection \cite{Tuy00}:
\begin{equation}
\theta^n  = \max\{\alpha\;|\;\alpha\boldsymbol{z}^n \in {\cal G}\},
 \end{equation}
where $\boldsymbol{z}^n = \arg \max_{\boldsymbol{z} \in {\cal T}_n} \; \Gamma(\boldsymbol{z})$ denotes the  maximizer among the vertices in ${\cal T}_n$ . With $\theta^n,$ the projection $\boldsymbol{y}^n =\theta^n \boldsymbol{z}^n $ is then the unique point where the halfline from 0 through $\boldsymbol{z}^n$ meets the upperboundary of ${\cal G}$. From (\ref{GG}), $\theta^n$ can be determined as
\begin{equation}\label{eq.mpmaxmin}
\begin{split}
\theta^n & = \max\{\alpha\;|\;\alpha\boldsymbol{z}^n \in {\cal G}\} \\
& = \max\{\alpha\;|\; \alpha \leq \min_{i=1,\ldots,2K} \frac{1+\text{SNR}_i(\boldsymbol{P})}{z^n_i}, \; \forall \boldsymbol{P} \in {\cal S}\} \\
& = \max_{\boldsymbol{P} \in {\cal S}} \; \min_{i=1,\ldots,2K}
     \frac{1+\text{SNR}_i(\boldsymbol{P})}{z^n_i}.
\end{split}
\end{equation}

With the definitions of $\text{SNR}_{BM,k}(\boldsymbol{P})$ in (\ref{eq.snrbm}) and $\text{SNR}_{MB,k}(\boldsymbol{P})$ in (\ref{eq.snrmb}), the
above max-min problem (\ref{eq.mpmaxmin}) can be decoupled into the following two sub-problems:
\begin{equation}\nonumber
\begin{split}
\text{P1:} ~~ & \theta_1^n  = \max_{\boldsymbol{P}_B} \;
\min_{k}
     \frac{1}{z^n_{k}} \left( 1+\left(\frac{|r_{BR}(k,k)|^2 P_{B,k} }{\sigma^2} -1 \right)^{+} \right) \\
& \text{s.t.} ~~ \sum_{k=1}^{K}  P_{B,k} \leq P_B.
\end{split}
\end{equation}

\begin{equation}\nonumber
\begin{split}
\text{P2:} ~~& \theta_2^n  = \max_{\boldsymbol{P}_R} \; \min \left(
 \min_{k}
     \frac{1}{z^n_{k}} \left( 1+ \frac{|l_{RM}(q_k,q_k)|^2 P_{R,k}}{\sigma^2} \right), \right. \nonumber \\
 & \quad \quad \quad \quad \quad \min_{k}
     \left. \frac{1 }{z^n_{K+k}} \left( 1+ \frac{|l_{RB}(q_k,q_k)|^2 P_{R,k}}{\sigma^2} \right) \right) \\
 & \text{s.t.} ~~ \sum_{k=1}^{K} P_{R,k} \leq P_R.
\end{split}
\end{equation}

The optimal $\theta^n$ for (\ref{eq.mpmaxmin}) is then given by $\theta^n = \min(\theta_1^n,\theta_2^n,\theta_3^n)$, where
\begin{equation}
\theta_3^n  = \min_{k=1,\ldots,K}
     \frac{1}{z^n_{K+k}} \left( 1 + \left(\frac{|r_{MR}(k,k)|^2 P_{M,k} }{\sigma^2} -1 \right)^{+} \right).
\end{equation}
Note that $ \theta_3^n$ simply follows from $\min_{i=K+1,\ldots,2K}
     \frac{1+\text{SNR}_i(\boldsymbol{P})}{z^n_i}$.

The solutions for  two sub-problems P1 and P2 are given in the following lemma.
 \lemma For P1, let ${\sigma}_i^2 := \frac{\sigma^2}{|{r}_{BR}(i,i)|^2},i=1,\ldots,K$, sort $z_i^n,i=1,\ldots,K,$ as
$z_{\pi_1}^n \leq z_{\pi_2}^n \leq \ldots \leq z_{\pi_{K}}^n $, and define $z_{\pi_0}^n=0$.  Then the optimal $\theta_1^n$ for problem P1 is given by
\begin{equation}\label{eq.p1.sol}
\theta_1^{n} = \max \left(\frac{1}{z_{\pi_{\ell}}^n},
\frac{P_B}{\sum_{k=\ell}^{K}  z_{\pi_k}^n {\sigma}_{\pi_k}^2}
\right),
\end{equation}
where  $\ell$ is an integer satisfying $1 \leq \ell \leq {K}$,
and $\frac{1}{z_{\pi_{\ell}}^n} \sum_{k=\ell+1}^{K}  z_{\pi_k}^n
{\sigma}_{\pi_k}^2 \leq P_B < \frac{1}{z_{\pi_{\ell-1}}^n}
\sum_{k=\ell}^{K}  z_{\pi_k}^n {\sigma}_{\pi_k}^2$.
For P2, the optimal $\theta_2^{n}$ is given by solving the following equation:
\begin{equation}\label{eq.opttheta2}
\sum_{k=1}^K \left[\max \left( \frac{(\theta_2^{n} z_k^n -1 ) \sigma^2}{|l_{RM}(q_k,q_k)|^2},\frac{(\theta_2^{n} z_{K+k}^n -1 ) \sigma^2}{|l_{RB}(q_k,q_k)|^2} \right) \right]^+ = P_R.
\end{equation}

\proof see Appendix B.

Let $\boldsymbol{z}^{\text{opt}}$ denote the global optimal solution of (\ref{eq.mp2}). For a given accuracy tolerance level $\epsilon>0$, we say that a
feasible $\bar{\boldsymbol{z}}$ is an $\epsilon$-optimal solution if
$(1+\epsilon) \Gamma(\bar{\boldsymbol{z}}) \geq
\Gamma(\boldsymbol{z}^{\text{opt}})$. Based on the max-min solution to (\ref{eq.mpmaxmin}), we propose the following algorithm to find an $\epsilon$-optimal solution to
(\ref{eq.mp2}).

\begin{quote}
\vspace{0.05 in}\hrule height0.1pt depth0.3pt \vspace{0.05 in}
\begin{algorithm}\label{algorithm.1}
{\it for weighted sum-rate maximization}

{\bf Initialize}: select an accuracy level $\epsilon>0$, let $n=0$,
 and $\text{CurrentBestValue(CBV)} = -\infty$. Initialize vertex set ${\cal T}_0$ with a selected outer vertex to construct the initial outer polyblock  ${\cal P}_0$.

{\bf Repeat}: \\
    1). Find $\boldsymbol{z}^n \in {\cal T}_n $ that maximizes $\Gamma(\boldsymbol{z})$ and solve (\ref{eq.mpmaxmin}) to obtain $\theta^n$ and $\boldsymbol{y}^n =  \theta^n \boldsymbol{z}^n$.\\
    2). If $\boldsymbol{y}^n \in {\cal H}$ and $\Gamma(\boldsymbol{y}^n) > \text{CBV}$, then $\text{CBV}=\Gamma(\boldsymbol{y}^n)$ and  $\bar{\boldsymbol{z}}=\boldsymbol{y}^n$. \\
    3).  Let $\boldsymbol{z}^{n}(i) = \boldsymbol{z}^n - (z^n_i - y^n_i) \boldsymbol{e}_i, i=1,\ldots,2K$, where $z^n_i$ and $y^n_i$ are the $i$-th entry of $\boldsymbol{z}^n$ and $\boldsymbol{y}^n$, respectively; i.e., $\boldsymbol{z}^{n}(i)$ is obtained by simply replacing the $i$-th entry of $\boldsymbol{z}^{n}$ by $y^n_i$. Clearly, $\boldsymbol{y}^n \le \boldsymbol{z}^n(i) \le \boldsymbol{z}^n$. \\
    4).  Let ${\cal T}_{n+1}=[({\cal T}_n \backslash \{\boldsymbol{z}^n\})\cup \{ \boldsymbol{z}^{k}(i)\}] \cap {\cal H}$; i.e., obtain a new vertex set ${\cal T}_{n+1}$ by replacing the vertex $\boldsymbol{z}^{n}$ in ${\cal T}_{n}$ with $2K$ new vertices $\boldsymbol{z}^n(i),i=1,\ldots,2K$. Further remove from ${\cal T}_{n+1}$  any $\boldsymbol{v}_j \in {\cal T}_{n+1}$  that satisfying $\Gamma(\boldsymbol{v}_j) \leq \text{CBV}(1 + \epsilon)$. By construction of $\boldsymbol{z}^n(i)$, the new outer polyblock ${\cal P}_{n+1}$ with vertex set ${\cal T}_{n+1}$ satisfies $ {\cal P}_{n+1} \subseteq {\cal P}_{n}$.\\
    5). Set $n=n+1$ and goto Step 1  until ${\cal T}_n$ is empty.

{\bf Output}: $\bar{\boldsymbol{z}}$ and CBV as the $\epsilon$-optimal solution for (\ref{eq.mp2}).
\end{algorithm}
\vspace{0.05 in} \hrule height0.1pt depth0.3pt \vspace{0.1 in}
\end{quote}

Per iteration of Algorithm 1, we have $\boldsymbol{y}^n = \theta^n \boldsymbol{z}^n  \in {\cal G}$. If $\boldsymbol{y}^n \in {\cal H}$ too, we obtain a feasible point $\boldsymbol{y}^n \in {\cal G} \cap {\cal H}$. In this case, we update $\text{CBV} = \max\{\text{CBV}, \Gamma(\boldsymbol{y}^n)\}$. This implies $\text{CBV}$ is the current best value so far, and the corresponding $\bar{\boldsymbol{z}}$ is the current best solution for (\ref{eq.mp2}). Observe that for any $\boldsymbol{v}_j \in {\cal T}_{n+1}$ satisfying $\Gamma(\boldsymbol{v}_j) \leq \text{CBV}(1 + \epsilon)$, we have $(1+\epsilon)\text{CBV} \geq \Gamma(\boldsymbol{y})$, $\forall \boldsymbol{y} \in [\boldsymbol{0}, \boldsymbol{v}_j]$, due to monotonicity of $\Gamma(\cdot)$. Hence, $\boldsymbol{v}_j$ can be removed from ${\cal T}_{n+1}$ for further consideration since $\bar{\boldsymbol{z}}$ will be the desired $\epsilon$-optimal solution if $\boldsymbol{z}^{\text{opt}} \in [\boldsymbol{0}, \boldsymbol{v}_j]$.

Algorithm 1 is essentially a ``smarter'' branch-and-bound method.  A key requirement for
its guaranteed convergence is that $\boldsymbol{z} \in {\cal G} \cap {\cal H}$ is
lower bounded by a strictly positive vector. Since $\boldsymbol{z}
\geq \boldsymbol{1} > \boldsymbol{0}$ in (\ref{eq.mp2}), it
readily follows from \cite[Theorem 1]{Tuy00} that
 \begin{proposition}
Algorithm 1 globally converges to an $\epsilon$-optimal solution for
(\ref{eq.mp2}).
\end{proposition}

\remark For each iteration of Algorithm 1, only a simple bisection search is involved (see Lemma 1). Furthermore, as shown in Theorem 2, equal power allocation is near optimal in many cases. Hence, we can use the equal power allocation solution to construct the initial outer polyblock ${\cal P}_0$, which can speed up the convergence of Algorithm 1.

Before leaving this section, we emphasize that the sum-rate optimization in this section works for a fixed DPC order at the relay, i.e., the permutation matrix $\boldsymbol{\Phi}$ in (\ref{eq.LQ}) is fixed. The global optimum will be obtained by enumerating over all possible DPC orders.

\section{Further Discussions}

\subsection{Extension to General Antenna Setups and Multiple Relays}
The proposed two-way relaying scheme can be readily extended to the MIMO cTWRC where each MS  is equipped with multiple antennas. In this case, a straightforward approach is to treat each antenna at the MS as a  virtual ``user" (though the virtual users associated with a common MS share the power budget of this MS). Then, the proposed transceiver and relaying scheme developed in Section III directly applies. Furthermore,  as antenna cooperation is physically allowed, we can introduce an extra linear precoder at each MS. The MSs' linear precoders as well as the power allocations at the BS and the RS can be designed using iterative optimization methods, such as the one in \cite{Fang13}. Closed-form suboptimal linear precoders can be also used, such as SVD-based eigen beamforming \cite{Yang13}. With an appropriate precoding design, the system performance can be further enhanced.

The proposed scheme can be extended to a more general antenna setup in which the number of antennas at the BS and the RS may be unequal. In this case, the total number of data streams that can be supported by the two-way relaying network is given by $N_{s}=2\min(N_B,N_R,\sum_{k=1}^{K}N_{M,k})$ with $N_{s}/2$ streams in each way, where $N_{M,k}$ represents the number of antennas at the $k$-th MS. When the number of antennas at the BS (or RS) is larger than $N_{s}/2$ (implying that this node has antenna redundancy in supporting $N_{s}/2$ independent data streams), an extra linear precoder can be applied at the BS (or RS) for beamforming. Again, iterative optimization methods could be used to find these linear precoders. The detailed precoding design for a general antenna setup will be an interesting direction for future research.

The proposed scheme can be also extended to the MIMO cTWRC with multiple distributed relay nodes \cite{RKW13}. One straightforward approach is to select the best relay for two-way communications based on  channel conditions. Then the proposed scheme directly applies. The system performance can be further improved if we jointly design encoding/decoding schemes at multiple relays. However, this is out of the scope of the present work and  will be a good direction to pursue in our future work.

\subsection{Channel Estimation in the MIMO cTWRC}
For the proposed scheme, global CSI is assumed available  at the BS and the RS for signal encoding/decoding and power allocation.  We make this assumption in order to explore the fundamental performance limit of  the cTWRC, as in the existing works [16]-[20]. From the viewpoint of practical implementation, CSI acquisition in two-way relay networks is  challenging. Generally,  the relay is deployed at fixed position, and the channel between the RS and BS is  slow-fading or  quasi-static. This simplifies the CSI estimation at
the BS. For the channels between the relay and the MSs, training based method can be employed to estimate the CSIs as in conventional cellular networks. These estimated CSIs at the RS can  then be fed back to the BS. Considering the significant performance improvements brought by the proposed scheme and  the fact that both BS and RS are powerful infrastructures, the complexity of the channel estimation in the MIMO cTWRC could be affordable.

\begin{figure}[t]
\centering
\includegraphics[width=3.3in]{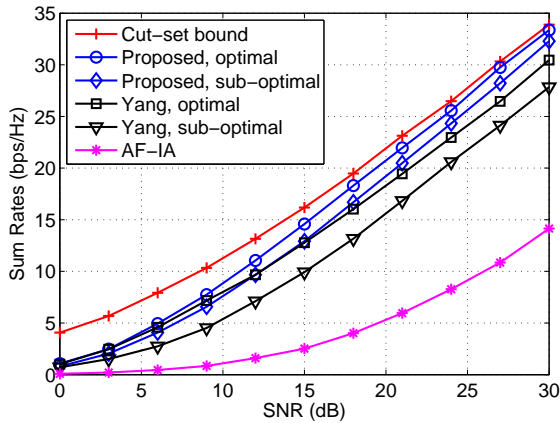}
\caption{Sum-rate performance of the cTWRC  with $K=4$ MSs, $P_B=P_R=P_M$.}
 \label{fig2}
\end{figure}

\begin{figure}[t]
\centering
\includegraphics[width=3.3in]{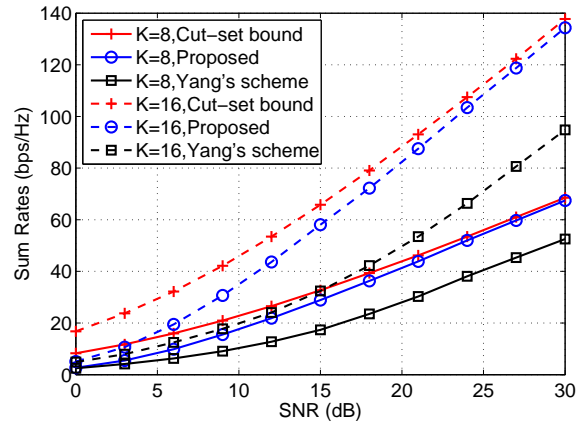}
\caption{Sum-rate performance of the cTWRC with large number of  MSs, $P_B=P_R=P_M$.}
 \label{fig3}
\end{figure}

\section{Numerical Results}

In this section, numerical results are presented to demonstrate the
performance of the proposed  two-way relaying scheme.  It
is assumed that all elements of $\boldsymbol{H}_{BR}$ and $\boldsymbol{h}_{k,R},k=1,\ldots,K $,  are independently drawn from a circularly  symmetric complex Gaussian distribution with zero mean and unit variance. The channel matrices/vectors remain constant over the
two phases and are reciprocal, i.e., $\boldsymbol{H}_{RB} = \boldsymbol{H}_{BR}^T$, $\boldsymbol{h}_{R,k} = \boldsymbol{h}_{k,R}^T, \forall k $. All the MSs have the same power budget, i.e., $P_{M,k}=P_M, \forall k$, and all the nodes have the same noise variance $\sigma^2$. The SNRs are defined as $SNR_{{\cal X},R} = P_{\cal X}/\sigma^2$ and $SNR_{R,{\cal X}} = P_R/\sigma^2$, where ${\cal X} \in \{B,M\}$. The $SNR$ in the following figures is defined as $SNR = SNR_{MR}= P_M/\sigma^2$.  Unless otherwise specified, it is also assumed that $N_M=1,N_B=N_R=K$, and each MS exchanges one data stream with the BS.

Fig. \ref{fig2} shows the sum-rate of the proposed scheme for a  cTWRC with $K=4$ single-antenna MSs. Each node in the network has the same power budget, i.e., $P_B=P_R=P_M$. Achieved sum-rate of the AF based interference alignment (AF-IA) scheme in \cite{Ding11} and the scheme proposed by Yang in \cite{YangHJ12} are included in Fig. \ref{fig2} for comparison. For the proposed scheme, we also consider a low-complexity sub-optimal alternative in which the DPC encoding order at the RS is fixed as $\boldsymbol{\Phi} = \boldsymbol{I}_K$ and equal power allocations are used for all spatial streams. A similar sub-optimal alternative for Yang's scheme is also shown, in which the BS uses ZF-based precoding, the RS employs fixed-order DPC and the transmit powers are the same for all data streams. It is shown that AF-IA scheme in \cite{Ding11} suffers from the noise propagation problem and the achievable sum-rate is much lower than the other two schemes. The proposed scheme with optimal permutation $\boldsymbol{\Phi}$ and power allocation can approach the cut-set bound as the SNR increase, the gap is only 0.5 bps/Hz when $SNR \ge 25$ dB. Compared with the sub-optimal scheme, there is about 1 bps/Hz gain for the optimal solution. It can be also  seen that sum-rate gain is about 3 bps/Hz for the proposed scheme as compared to Yang's scheme.

Fig. \ref{fig3} compares the sum-rate performance of the proposed scheme and the Yang's scheme \cite{YangHJ12} when there are $K=8$ and $K=16$ single-antenna MSs in the cTWRC. Note that there are $K!$ possible DPC encoding orders in total at the RS. Even for a moderate $K$, it is too time-consuming to find the optimal DPC encoding order and the corresponding optimal power allocations. Instead, we search over $5 \times 10^4$ randomly chosen DPC encoding orders and use equal power allocation to determine the best DPC order. Even with this suboptimal approach, we can see from Fig. 3 that the proposed scheme performs only 0.5 dB away from the cut-set bound. Compared with Yang's scheme, we see that the performance gain increases as the number of MSs increases. In particular, when there are 16 users in the network, more than 7 dB power gain is observed for the proposed scheme.

\begin{figure}[t]
\centering
\includegraphics[width=3.3in]{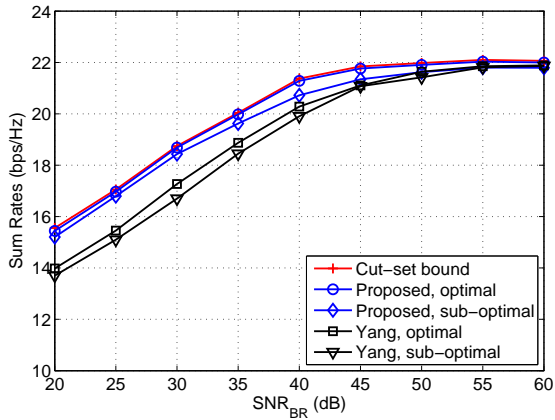}
\caption{Sum-rate vs. $SNR_{BR}$ for the cTWRC with $K=2$  MSs, $SNR_{MR}=30$ dB, and $SNR_{RB}=SNR_{RM}=40$ dB.}
 \label{fig.simupb}
\end{figure}

To further demonstrate the advantage of the proposed scheme, Fig. \ref{fig.simupb} illustrates the achievability of the cut-set bound under different $SNR_{BR}$. The transmit power of the RS is 10 dB higher than the MSs, and the SNRs are fixed to be $SNR_{MR} = 30$ dB, $SNR_{RM}=SNR_{RB}=40$ dB. It can be seen that the proposed scheme is able to achieve the cut-set bound when $SNR_{BR} \ge 20$ dB. However, for Yang's scheme, it is required that $SNR_{BR}$ is much larger than $SNR_{RM}$ to meet the condition for the rate-loss due to precoding at the
BS to be negligible.

\begin{figure}[t]
\centering
\includegraphics[width=3.3in]{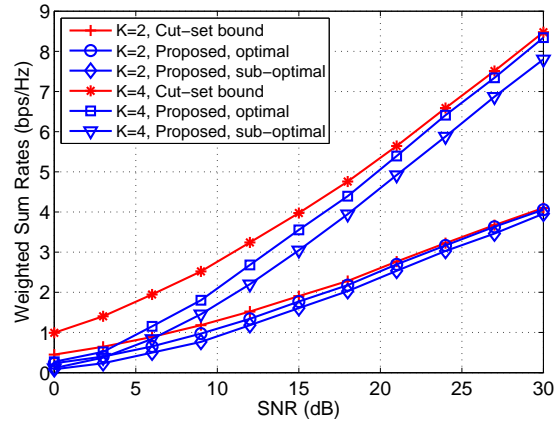}
\caption{Weighted sum-rate performance of the cTWRC , $P_B=P_R=P_M$, $\xi_{B,k}=0.4$, and $\xi_{M,k}=0.1, \forall k$.}
 \label{fig.unequalweight}
\end{figure}

The weighted sum-rate performance of the proposed scheme is shown in Fig. \ref{fig.unequalweight}. We assume that the priority of the data transmission from the BS to the MSs are higher than that from the MSs to the BS. The weights are chosen as $\xi_{B,k}=0.4$ and $\xi_{M,k}=0.1, \forall k$.  Again, the proposed scheme is able to approach the weighted sum-rate cut-set bound when the SNR is higher than 25 dB.

\begin{figure}[t]
\centering
\includegraphics[width=3.3in]{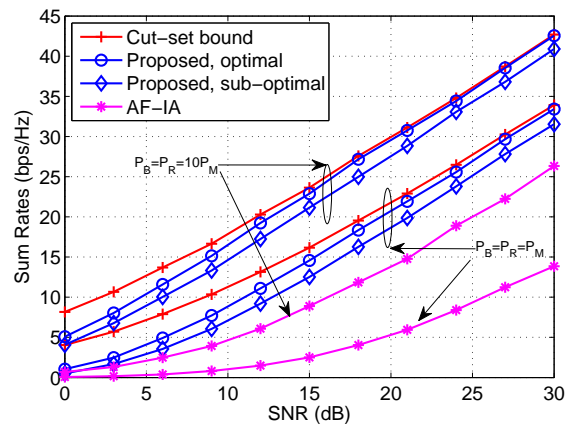}
\caption{Performance of the MIMO cTWRC with $K=2$ multi-antenna MSs. Each MS is equipped with $N_M=2$ antennas, and $N_B=N_R=4$. }
 \label{fig.mimotwrc}
\end{figure}

Finally, Fig. \ref{fig.mimotwrc} shows the performance of the MIMO cTWRC with $K=2$ multi-antenna MSs. The number of antennas at the BS and RS is set to $N_B=N_R=4$ and each MS is equipped with $N_M=2$ antennas. Each MS exchanges two data streams with the BS; at any of the two MSs, no extra precoding is applied, i.e., each antenna transmit an independent data stream. Again, we see from Fig. \ref{fig.mimotwrc} that the sum-rate upper bound of the MIMO cTWRC can be asymptotically achieved within 0.5 bps/Hz.

\section{Conclusion}
We proposed a novel  two-way DF relaying scheme for the MIMO cTWRC.  A non-linear lattice precoder was proposed at the BS to pre-compensate for the inter-stream interference, which enables efficient interference-free lattice decoding at the relay. The sufficient conditions for the achievability of the sum-rate cut-set bound were derived. The optimal power allocation for weighted sum-rate maximization was also obtained through monotonic programming. Numerical results demonstrated that the proposed scheme outperforms the existing alternatives and closely approaches the cut-set bound in the high SNR regime.

%
%
\section*{Appendix A: Proof of Theorem 2}

Consider the proposed scheme with equal power allocation, i.e., $P_{B,k}=P_B/K$ and $P_{R,k}=P_R/K, \forall k$. In the high SNR regime, the achievable rates $ R_{B \to R,k}(P_{B,k})$ and $ R_{R \to M,k}(P_{R,k})$ can be expressed as
\begin{equation}
\begin{split}
 R_{B \to R,k}(P_{B,k}) &= \left[ \frac{1}{2} \log\left( \frac{|r_{BR}(k,k)|^2 P_{B,k} }{\sigma^2}\right) \right]^+ \\
 &\simeq \frac{1}{2} \log\left( \frac{|r_{BR}(k,k)|^2 P_{B,k} }{\sigma^2}\right),
 \end{split}
\end{equation}
and
\begin{equation}
R_{R \to M,k}(P_{R,k})  \simeq \frac{1}{2} \log\left( \frac{|l_{RM}(q_k,q_k)|^2 P_{R,k} }{\sigma^2}\right).
\end{equation}

We first consider Condition C1. If the inequalities in C1 holds, i.e,
\begin{equation}
P_B \le \frac{ |l_{RM}(q_k,q_k)|^2}{|r_{BR}(k,k)|^2}  P_R,
\end{equation}
then we have
\begin{eqnarray}
R_{B \to R,k}(P_{B,k}) &=&  \frac{1}{2} \log \left( \frac{|r_{BR}(k,k)|^2 P_B}{K\sigma^2} \right)
\nonumber \\
&\le&  R_{R \to M,k}(P_{R,k}),
\end{eqnarray}
i.e., the achievable rate for the $k$-stream of the BS-RS link is lower than or equal to that from the RS to the $k$-th MS. Consequently, the sum-rate of the proposed scheme with equal power allocation  for the BS-to-MS link is given by
\begin{eqnarray}\label{eq.a.prop}
R_{\text{sum}, B \to M} &=&  \sum_{k=1}^{K} R_{B \to R,k}(P_{B,k})  \nonumber \\
&=&  \sum_{k=1}^{K} \frac{1}{2} \log \left( \frac{|r_{BR}(k,k)|^2 P_B }{K \sigma^2} \right) \nonumber \\
&=&   \frac{1}{2} \log \prod_{k=1}^{K}  \left( \frac{|r_{BR}(k,k)|^2 P_B }{K \sigma^2} \right) \nonumber\\
& =&  \frac{1}{2} \log \det \left( \frac{P_B}{K \sigma^2} \boldsymbol{H}_{BR} \boldsymbol{H}_{BR}^H \right) \nonumber \\
& = & \sum_{k=1}^{K}  \frac{1}{2} \log \left( \frac{\lambda_{BR,k} P_B} {K \sigma^2} \right)
\end{eqnarray}

Now consider the cut-set bound in (\ref{eq.cutset}). Note that C1 implies $P_B \le \rho_{B}  P_R$, then the sum-rate bound of the BS-to-MS link is given by
\begin{equation}\label{eq.a.cutset}
 R_{\text{sum,cs},B \to M} =  \sum_{k=1}^{K}  R_{B,k} = \sum_{k=1}^{K}  \frac{1}{2} \log \left( \frac{\lambda_{BR,k} P_B} {K \sigma^2} \right).
 \end{equation}

From (\ref{eq.a.prop}) and (\ref{eq.a.cutset}), we see that the gap between the proposed scheme and the cut-set upper bound of the BS-to-MS link vanishes as $\frac{P_B}{\sigma^2}  \to +\infty$. Therefore, the proposed scheme asymptotically achieves the cut-set upper bound of the BS-to-MS link when condition C1 holds. Similarly, it can be shown that when C2 holds, the proposed scheme asymptotically achieves the cut-set upper bound of the BS-to-MS link as $\frac{P_R}{\sigma^2}  \to +\infty$.

Further,  it can be shown in a similar way that the sum-rate gap between the proposed scheme and the cut-set bound for the MS-to-BS link vanishes as $\frac{P_{M,k}}{\sigma^2} \to +\infty$ and $\frac{P_R}{\sigma^2} \to +\infty$ when either C3 or C4 holds. This concludes the proof of Theorem 2.

 \section*{Appendix B: Proof of Lemma 1}

For  P1, it is clear that $\sum_{k=1}^K P_{B,k} = P_B$ holds for the optimal $\theta_1$. Then P1 can be written as:
\begin{subequations}\label{eq.Amaxmin}
\begin{align}
& \theta_1^n  = \max_{\boldsymbol{P}_B} \; \min_{k=1,\ldots,K}
     \frac{1+ \left(\frac{P_{B,k}}{\sigma_k^2} -1 \right)^{+}}{z_k^n} \\
 & \text{s.t.}~~\sum_{k=1}^K   P_{B,k} =  P_B,
\end{align}
\end{subequations}
where $\sigma_k^2 = \sigma^2/|r_{BR}(k,k)|^2$.

Sort $\{z_k^n\}_{k=1}^K$ that
$z_{\pi_1}^n \leq z_{\pi_2}^n \leq \ldots \leq z_{\pi_K}^n $ and define $z_{\pi_0}^n =0$. Note that $\theta_1^n \geq \frac{1}{z_{\pi_K}^n}$. Depend on the value of $P_B$, there are two possible cases:

 {\it Case A:}  $\theta_1^n = \frac{1}{z_{\pi_{\ell}}^n} $,  where $1 \leq \ell \leq K$. The corresponding power allocation satisfies
\begin{subequations}
\begin{align}
& P_{B,\pi_k} =0,  \quad \frac{1 \!+\!
\left(\frac{P_{B,\pi_k}}{\sigma_{\pi_k}^2} \! - \! 1
\right)^{+}}{z_{\pi_k}^n} = \frac{1}{z_{\pi_k}^n}, \quad k\!\!=\!\!1,  \ldots  ,\ell \!-\! 1, \\
& 0 \leq P_{B,\pi_\ell} \leq \sigma_{\pi_\ell}^2, \quad \frac{1+
\left(\frac{P_{B,\pi_\ell}}{\sigma_{\pi_\ell}^2} -1
\right)^{+}}{z_{\pi_\ell}^n} = \frac{1}{z_{\pi_{\ell}}^n}, \\
& P_{B,\pi_k} > \sigma_{\pi_k}^2, \quad \frac{1 \!+\!
\left(\frac{P_{B,\pi_k}}{\sigma_{\pi_k}^2} \!-\! 1
\right)^{+}}{z_{\pi_k}^n} = \frac{1}{z_{\pi_{\ell}}^n},  \quad k \!=\! \ell \!+\! 1,\ldots, K. \label{eq.b.a.b}
\end{align}
\end{subequations}
From (\ref{eq.b.a.b}), we have $P_{B,\pi_k} = \frac{z_{\pi_k}^n \sigma_{\pi_k}^2}{z_{\pi_{\ell}}}$, for $ k= \ell+1,\ldots,K$. Then, $P_B = \sum_{k=1}^K   P_{B,k} = \sum_{k=\ell+1}^K
P_{B,\pi_k} + P_{B,\pi_\ell}$, and we have
\begin{equation}\label{eq.PB21}
\frac{1}{z_{\pi_{\ell}}^n} \sum_{k=\ell+1}^K z_{\pi_k}^n
\sigma_{\pi_k}^2 \leq P_B \leq \frac{1}{z_{\pi_{\ell}}^n}
\sum_{k=\ell}^K  z_{\pi_k}^n \sigma_{\pi_k}^2.
\end{equation}

{\it Case B:} $\frac{1}{z_{\pi_{\ell}}^n} < \theta_1^n <
\frac{1}{z_{\pi_{\ell-1}}^n}$,  where $1 \leq \ell \leq K$. The corresponding power
allocation satisfies
\begin{subequations}
\begin{align}
& P_{B,\pi_k} =0,  \quad \frac{1+
\left(\frac{P_{B,\pi_k}}{\sigma_{\pi_k}^2} -1
\right)^{+}}{z_{\pi_k}^n} = \frac{1}{z_{\pi_k}^n}, \quad k=1,\ldots,\ell-1, \\
& P_{B,\pi_k} > \sigma_{\pi_k}^2, \quad \frac{1+
\left(\frac{P_{B,\pi_k}}{\sigma_{\pi_k}^2} -1
\right)^{+}}{z_{\pi_k}^n} = \theta_1^n , \quad k= \ell,\ldots,K. \label{eq.a.caseb.b}
\end{align}
\end{subequations}
From (\ref{eq.a.caseb.b}), we have $P_{B,\pi_k} =  z_{\pi_k}^n \sigma_{\pi_k}^2
\theta_1^n$, for $k= \ell,\ldots,K$. Then, $P_B = \sum_{k=1}^K   P_{B,k} = \sum_{k=\ell}^K
P_{B,\pi_k} = \sum_{k=\ell}^K  z_{\pi_k}^n \sigma_{\pi_k}^2
\theta_1^n $, and $\theta_1^n =
\frac{P_B}{\sum_{k=\ell}^K   z_{\pi_k}^n \sigma_{\pi_k}^2}$. From  $\frac{1}{z_{\pi_{\ell}}^n} < \theta_1^n <
\frac{1}{z_{\pi_{\ell-1}}^n}$, we obtain the constrains on $P_B$ as
\begin{equation}\label{eq.PB2}
\frac{1}{z_{\pi_{\ell}}^n} \sum_{k=\ell}^K  z_{\pi_k}^n
\sigma_{\pi_k}^2 < P_B < \frac{1}{z_{\pi_{\ell-1}}^n}
\sum_{k=\ell}^K  z_{\pi_k}^n \sigma_{\pi_k}^2.
\end{equation}

Combining the above two cases we arrive at (\ref{eq.p1.sol}).

For P2, the optimal solution must satisfy
\begin{equation}
\frac{1}{z^n_{k}} \left( 1+ \frac{|l_{RM}(q_k,q_k)|^2 P_{R,k}}{\sigma^2} \right)\ge \theta_2^n,
\end{equation}
which implies that $ P_{R,k} \ge \frac{(\theta_2^{n} z_k^n -1 ) \sigma^2}{|l_{RM}(q_k,q_k)|^2}$. Similarly, we have $P_{R,k} \ge \frac{(\theta_2^{n} z_{K+k}^n -1 ) \sigma^2}{|l_{RB}(q_k,q_k)|^2}$. On the other hand, $P_{R,k} \ge 0$, hence we have
\begin{equation}
P_{R,k} \ge \left[ \max \left( \frac{(\theta_2^{n} z_k^n -1 ) \sigma^2}{|l_{RM}(q_k,q_k)|^2}, \frac{(\theta_2^{n} z_{K+k}^n -1 ) \sigma^2}{|l_{RB}(q_k,q_k)|^2} \right) \right]^+.
\end{equation}
As a result, the maximum  $\theta_2^{n}$ can be determined by solving the equation (\ref{eq.opttheta2}) via a simple bisection search.



\begin{thebibliography}{99}

\bibitem{Shannon61} C. E. Shannon, ``Two-way communication channels,'' in {\em Proc. 4th
Berkeley Symp. Math. Stat. Prob.}, Berkely, CA,  vol. 1, pp. 611-644, 1961.
\bibitem{Zhang06} S. Zhang, S.-C. Liew, and P. P. Lam, ``Hot topic: Physical-layer network coding,'' in {\em Proc. ACM MobiCom}, Los Angeles, CA, 2006.
\bibitem{Ran07} B. Rankov and A. Wittneben, ``Spectral efficient protocols for half-duplex fading relay channels,'' {\em IEEE J. Sel. Areas Commun.}, vol. 25, no.
2, pp. 379-389, Feb. 2007.
\bibitem {Pop07} P. Popovski and H. Yomo, ``Physical network coding in two-way wireless
relay channels,'' in {\em Proc. ICC}, Glasgow, U.K., June 2007.

\bibitem{Naz11} B. Nazer and M. Gasper, ``Compute-and-forward: Harnessing interference through structured
codes,'' {\em IEEE Trans. Inf. Theory}, vol. 57, no. 10, pp. 6463--6486, Oct. 2011.

\bibitem{Nam10} W. Nam, S. Chung, and Y. H. Lee, ``Capacity of the Gaussian two-way
relay channel to within 1/2 bit,'' {\em IEEE Trans. Inf. Theory}, vol. 56, no.
11, pp. 5488-5494, Nov. 2010.

\bibitem{Yang13} T. Yang, X. Yuan, P. Li, I. B. Collings, and J. Yuan, ``A new physical-layer network coding scheme with eigen-direction alignment precoding for MIMO two-way relaying,'' {\em IEEE Trans. Commun.}, vol. 61, no. 3, pp. 973-986, March 2013.

\bibitem{Yuan13} X. Yuan, T. Yang, and I. B. Collings, ``Multiple-input multiple-output two-way relaying: A space-division approach ,'' {\em IEEE Trans. Inf. Theory}, vol. 59, no. 10, pp. 6421-6440, Oct. 2013.
\bibitem{Yang11}   H. J. Yang, J. Chun, and A. Paulraj, ``Asymptotic capacity of the separated MIMO two-way relay channel,'' {\em IEEE Trans. Inf. Theory}, vol. 57, no. 11, pp. 7542-7554, Nov. 2011.


\bibitem{Gun13} D. Gunduz, A. Yener, A. Goldsmith, and H. V. Poor, ``The multiway relay channel,'' {\em  IEEE Trans. Inf. Theory}, vol. 59, no. 1, pp. 51-63, Jan. 2013.

 \bibitem{Che09} M. Chen and A. Yener, ``Multiuser two-way relaying: Detection and interference management strategies,'' {\em IEEE Trans. Wireless Commun.}, vol. 8, pp. 4296-4303, Aug. 2009.

\bibitem{Tao12} M. Tao and R. Wang, ``Linear precoding for multi-pair two-way MIMO relay systems with max-min fairness,'' {\em IEEE Trans. Signal Process.}, vol. 60, no. 10, pp. 5361--5370, Oct. 2012.

\bibitem{Zha12}  J. Zhang, F. Roemer, and M. Haardt, ``Relay assisted physical resource
sharing: Projection based separation of multiple operators (ProBaSeMO)
for two-way relaying with MIMO amplify and forward relays,'' {\em IEEE
Trans. Signal Process}., vol. 60, no. 9, pp. 4834-4848, Sept. 2012.

\bibitem{Fang13} Z. Fang, X. Wang, and X. Yuan, ``Beamforming design for multiuser two-way relaying: A unified approach via max-min SINR,'' {\em IEEE Trans. Signal Process.}, vol. 61, no. 23, pp. 5841-5852, Dec. 2013.



\bibitem{Lee10} N. Lee, J.  Lim, and J. Chun, ``Degrees of freedom of the MIMO Y channel: Signal space alignment for network coding,'' {\em IEEE Trans. Inf. Theory}, vol. 56, no. 7, pp. 3332-3342, July 2010.



\bibitem{Bhat12} P. Bhat,  S. Nagata,  L. Campoy, et al., ``LTE-Advanced: An operator perspective,'' {\em IEEE Commun. Mag.}, vol. 50, no. 2, pp. 104-114, Feb. 2012.


\bibitem{Ding11} Z. Ding, I. Krikidis, J. Thompson, and K. K. Leung, ``Physical layer network coding and precoding for the two-way relay
channel in cellular systems,'' {\em IEEE Trans. Signal Process.}, vol. 59, no. 2, pp. 696-712, 2011.

\bibitem{Sun12} C. Sun, C. Yang, Y. Li, and B. Vucetic, ``Transceiver design for multiuser multi-antenna two-way relay cellular systems,'' {\em  IEEE
Trans. Commun.}, vol. 60, no. 10, pp.2893-2903, Oct. 2012.

\bibitem{Chi12} E. Chiu and V. K. N. Lau, ``Cellular multiuser two-way MIMO AF relaying via signal space alignment: Minimum weighted
SINR maximization,'' {\em IEEE Trans. Signal Process}., vol. 60, no. 9, pp. 4864-4873, Sept. 2012.

\bibitem{Gan13} M. Gan, Z. Ding, and X. Dai, ``Application of analog network coding to MIMO two-way relay channel in cellular systems,'' {\em IEEE Trans. Signal Process}.,  vol. 20, no.7, pp. 641-644, July 2013.

\bibitem{YangHJ12} H. J. Yang, Y. Choi, N. Lee, and A. Paulraj, ``Achievable sum-rate of MU-MIMO cellular two-way relay channels: Lattice code-aided linear precoding'', {\em IEEE J. Sel. Areas Commun.}, vol. 3, no. 8, pp. 1304-1318, 2012.

\bibitem{Wan13} F. Wang, X. Yuan, S. C. Liew, and Y. Li, ``Bidirectional cellular relay network with distributed relaying,'' {\em  IEEE J. Sel. Areas Commun.}, vol. 31, no. 10, pp. 2082-2098, Oct. 2013.




\bibitem{Tuy00} H. Tuy, ``Monotonic optimization: Problems and solution approaches,'' {\em SIAM J. Optim.}, vol. 11, no. 2, pp. 464-494, 2000.


\bibitem{Zam02}  R. Zamir, S. Shamai, and U. Erez, ``Nested linear/lattice codes for
structured multiterminal binning,'' {\em IEEE Trans. Inf. Theory}, vol. 48, no. 6, pp. 1250-1276, June 2002.

\bibitem{Erez04}  U. Erez and R. Zamir, ``Achieving 1/2 log(1 + SNR) on the AWGN
channel with lattice encoding and decoding,'' {\em IEEE Trans. Inf. Theory},
vol. 50, no. 10, pp. 2293-2314, Oct. 2004.

\bibitem{Erez05} U. Erez, S. Litsyn, and R. Zamir, ``Lattices which are good for (almost)
everything,'' {\em IEEE Trans. Inf. Theory}, vol. 51, no. 10, pp. 3401-3416, Oct. 2005.

\bibitem{Cai03}   G. Caire and S. Shamai, ``On the achievable throughput of a multiantenna Gaussian broadcast channel,'' {\em IEEE Trans. Inf. Theory}, vol. 49, no. 7, pp. 1691-1706, July 2003.

\bibitem{Qian09} L. P. Qian, Y. J. Zhang, and J. Huang, ``MAPEL: Achieving global
optimality for a non-convex wireless power control problem,'' {\em IEEE
Trans. Wireless Commun.}, vol. 8, no. 3, pp. 1553-1563, Mar. 2009.

\bibitem{Uts12} W. Utschick and J. Brehmer, ``Monotonic optimization framework for
coordinated beamforming in multicell networks,'' {\em IEEE Trans. Signal
Process.}, vol. 60, no. 4, pp. 1899-1909, 2012.

\bibitem{RKW13}  R. Rolny, M. Kuhn, and A. Wittneben, ``The relay carpet:
Ubiquitous two-way relaying in cooperative cellular networks,''
in {\em Proc. PIMRC }, pp. 1174-1179, Sept. 2013.


\end{thebibliography}
\end{document}